\documentclass[aps,12pt,superscriptaddress,floatfix,nofootinbib,altaffilletter]{revtex4-1}
 \usepackage{graphicx}
\usepackage[colorlinks=true,citecolor=darkred,urlcolor=darkred, pdfborder={0 0 0}]{hyperref}
\usepackage[utf8]{inputenc} 
\usepackage{amsmath,amssymb,amsbsy}
\usepackage[usenames,dvipsnames]{color} 
\usepackage{float}
\usepackage{calc}
\usepackage{comment}
\usepackage{multirow}
\usepackage{slashed}
\usepackage[normalem]{ulem}
\usepackage{braket}
\usepackage{mathtools}
\usepackage{caption}
\usepackage{subfigure}
\usepackage{wrapfig}
\reversemarginpar
\usepackage{xcolor}

\usepackage{paralist}
\usepackage{comment}
\usepackage[T1]{fontenc}
\usepackage{enumitem}
\usepackage{cancel}
\usepackage{ragged2e}

\definecolor{linkcolor}{rgb}{0,0,0.5}
\definecolor{darkred}{rgb}{0.6,0,0}
\def\bl#1{\textcolor{blue}{#1}}

\def\black{\color{black}{}}

\newcommand {\ignore}[1]{}

\def\lnv{lepton number violation }
\def \znbb {$0\nu\beta\beta$ }

\newcommand{\AddrAHEP}{
  AHEP Group, Institut de F\'{i}sica Corpuscular --
  CSIC/Universitat de Val\`{e}ncia, Parc Cient\'ific de Paterna.\\
 C/ Catedr\'atico Jos\'e Beltr\'an, 2 E-46980 Paterna (Valencia) - SPAIN}
 \newcommand{\AddrSOTON}{Department of Physics and Astronomy, University of Southampton, SO17 1BJ Southampton, United Kingdom}
 \newcommand{\AddrUTFSM}{Universidad T\'{e}cnica Federico Santa Mar\'{\i}a, Casilla 110-V, Valpara\'{\i}so, Chile.}
 \newcommand{\AddrCCTVAL}{{Centro Cient\'{\i}fico-Tecnol\'ogico de Valpara\'{\i}so, Casilla 110-V,
Valpara\'{\i}so, Chile}}
 \newcommand{\AddrSAPHIR}{{Millennium Institute for Subatomic Physics at High-Energy Frontier
(SAPHIR), Fern\'andez Concha 700, Santiago, Chile}}
\newcommand{\AddrUNAB}{Departamento de Ciencias F\'isicas, Universidad Andr\'es Bello, Sazi\'e 2212, Piso 7, Santiago, Chile}

\begin{document}

\title{\color{BrickRed} 
 TeV-Scale Leptogenesis in the Linear Seesaw Framework}

\author{A. E. C\'arcamo Hern\'andez\,}\email{antonio.carcamo@usm.cl}
\affiliation{\AddrUTFSM}
\affiliation{\AddrCCTVAL}
\affiliation{\AddrSAPHIR}
\author{J. Echeverria-Puentes}\email{j.echeverriapuentes@uandresbello.edu}
\affiliation{\AddrUNAB}
\author{Stephen F. King}
\email{king@soton.ac.uk}
\affiliation{\AddrSOTON}
\author{Jos\'{e} W. F. Valle}\email{valle@ific.uv.es}
\affiliation{\AddrAHEP}

\begin{abstract} 
The linear seesaw mechanism relates the smallness of neutrino masses to the mass splitting within a quasi-Dirac pair of heavy neutrinos, thereby providing a natural framework for realising resonant leptogenesis at low energies. We investigate this possibility within its simplest effective realisation, in which neutrino masses are mediated by a quasi-Dirac pair at the TeV scale. We estimate the generation of the baryon asymmetry in the purely flavoured and strong-washout regimes, identifying phenomenologically viable parameter regions and benchmark scenarios in which the heavy-neutrino decays are dominated either by the SM Higgs doublet or by a second, leptophilic scalar doublet responsible for neutrino mass generation. We find viable leptogenesis solutions both near and away from the resonant maximum and discuss complementary phenomenological probes of the framework.
\end{abstract}
\maketitle
\section{Introduction}\label{sec:introduction}

Leptogenesis is the mechanism whereby a lepton asymmetry is generated in the early universe 
and subsequently converted into the observed cosmological baryon asymmetry by sphaleron processes. This picture is naturally realised in high-scale seesaw schemes of neutrino mass generation~\cite{Fukugita:1986hr,Buchmuller:2005eh,Buchmuller:2004nz,Davidson:2008bu}.
An interesting scenario arises when a pair of Majorana neutrinos has a mass difference comparable with its decay widths. 
In this regime, the CP asymmetry is dominated by the self-energy effects and is resonantly amplified, allowing baryon asymmetry generation at smaller energy scales, a scenario known as Resonant Leptogenesis \cite{Pilaftsis:2003gt,Pilaftsis:2005rv,Deppisch:2010fr,BhupalDev:2014pfm,CarcamoHernandez:2024edi,King:2026hnn}.

The minimal linear seesaw mechanism~\cite{Akhmedov:1995ip,Akhmedov:1995vm,Malinsky:2005bi} is a compelling Standard Model (SM) extension that explains the smallness of neutrino masses without requiring the extremely high energy scales of the traditional seesaw mechanism. It achieves this by introducing new fermions and scalars supplemented by a lepton number symmetry.

Here we develop a predictive framework for low-scale leptogenesis based on the minimal missing partner version~\cite{Schechter:1980gr} of the linear seesaw mechanism~\cite{Batra:2022arl,Batra:2023ssq,Batra:2023mds}. Our analysis focuses on its minimal effective realisation, in terms of the Standard Model gauge group, explicit lepton-number violation, and a single quasi-Dirac neutrino mediator.
When lepton number symmetry is explicitly and softly broken, the associated VEV is naturally proportional to the soft-breaking mass parameter, and neutrino masses are symmetry protected. Since this VEV controls both the active neutrino mass scale and the mass splitting of the heavy quasi-Dirac mediator pair, the smallness of neutrino masses can naturally lead to a regime in which the heavy-neutrino decay widths are comparable to their mass splitting. This opens up the possibility of resonant enhancement of the CP asymmetry and, consequently, of successful Resonant Leptogenesis.

A more complete dynamical scenario with spontaneous  lepton number violation~\cite{Chikashige:1980ui,Schechter:1981cv} was considered in~\cite{Fontes:2019uld} and could also be adopted as an interesting framework for resonant leptogenesis. However we will focus here on the minimal effective scenario, as it captures the essential features, namely, it naturally incorporates the mechanism of resonant leptogenesis proposed in~\cite{Pilaftsis:2003gt} and allows for tighter phenomenological predictions. 
Indeed, our minimal construction 
allows for a direct connection between the physics responsible for the matter-antimatter asymmetry and the parameters we can measure in current neutrino oscillation
experiments as well as upcoming studies.

A similar rationale has been explored in the context of Type-Ib seesaw schemes~\cite{Hernandez-Garcia:2019uof,Chianese:2021toe,Fu:2021fyk}. Although the philosophy of the two approaches is quite different, there are common features and also important differences, making them complementary. 
Resonant leptogenesis has also been considered in several inverse seesaw scenarios~\cite{Blanchet:2009kk,Gu:2010xc,Blanchet:2010kw,Aoki:2015owa,Abada:2015rta,Abada:2017ieq,Agashe:2018cuf,Dolan:2018yqy,Hernandez:2021kju,Hernandez:2021uxx,Hernandez:2021xet,Chakraborty:2021azg,Abada:2023zbb,Mukherjee:2023nyi,Binh:2024lez,Abada:2026ons}. 

\section{Linear seesaw mechanism}
\label{sec:model}

We consider the minimal effective realisation of the linear seesaw mechanism~\cite{Akhmedov:1995ip,Akhmedov:1995vm,Malinsky:2005bi} in terms of the standard $SU(3)_C\otimes SU(2)_L\otimes U(1)_Y$ gauge symmetry, supplemented by an explicitly broken lepton-number symmetry. In addition to the SM Higgs doublet $\Phi$, which carries no lepton number, the model contains one scalar doublet carrying lepton number, denoted by $\chi_L$, and a single pair of mediator fermion singlets, $\nu^c$ and $S$, with opposite lepton-number assignments~\cite{Batra:2022arl,Batra:2023ssq,Batra:2023mds}: 
\begin{equation}
\begin{aligned}
L[L_\alpha]=1,\qquad
L[\Phi]=0,\qquad
L[\chi_L]=-2,\qquad
L[\nu^c]=-1,\qquad
L[S]=1,
\end{aligned}
\end{equation}
where $\alpha=e,\mu,\tau$ denotes the lepton flavour. The interactions of these fields are responsible for the generation of active neutrino masses. The Yukawa Lagrangian relevant for neutrino mass generation is given by
\begin{equation}
\label{eq:yukawa_lagrangian}
\begin{aligned}
-\mathcal{L}_{Y}
=
Y_\nu^\alpha L_\alpha^T C \nu^c \Phi
+
M_R \nu^c C S
+
Y_S^\alpha L_\alpha^T C S \chi_L
+
\mathrm{h.c.}
\end{aligned}
\end{equation}
Here $\mathbf{Y}_\nu$ and $\mathbf{Y}_S$ are vectors in flavour space, while $M_R$ is a Dirac-like mass term for the singlet mediators. It is important to note that this sector conserves lepton number. 

Once electroweak symmetry breaking takes place, the neutral components of the scalar doublets acquire vacuum expectation values
\begin{equation}
\label{eq:scalar_doublets}
\begin{aligned}
\langle \Phi^0\rangle=\frac{v_\Phi}{\sqrt{2}},
\qquad
\langle \chi_L^0\rangle=\frac{v_\chi}{\sqrt{2}},
\qquad
v^2=v_\Phi^2+v_\chi^2.
\end{aligned}
\end{equation}
In the basis $(\nu,\nu^c,S)$, the neutral fermion mass matrix takes the form 
\begin{equation}
\label{eq:full_neutrino_matrix}
\begin{aligned}
\mathcal{M}_\nu=
\begin{pmatrix}
0 & \mathbf{m}_D & \mathbf{M}_L \\
\mathbf{m}_D^T & 0 & M_R \\
\mathbf{M}_L^T & M_R & 0
\end{pmatrix},
\end{aligned}
\end{equation}
where the corresponding Dirac-like mass entries form vectors are given by
\begin{equation}
\begin{aligned}
\mathbf{m}_D=\frac{\mathbf{Y}_\nu v_\Phi}{\sqrt{2}},
\qquad
\mathbf{M}_L=\frac{\mathbf{Y}_S v_\chi}{\sqrt{2}}.
\end{aligned}
\end{equation} 
In the seesaw regime $M_R\gg |\mathbf{m}_D| \gg |\mathbf{M}_L|$, the effective light-neutrino mass matrix depends linearly on $\mathbf{m}_D$ and is given by 
\begin{equation}
\begin{aligned}
\mathbf{m}_\nu
=
\frac{\mathbf{m}_D \mathbf{M}_L^T+\mathbf{M}_L \mathbf{m}_D^T}{M_R}
=
\frac{v_\Phi v_\chi}{2M_R}
\left(
\mathbf{Y}_\nu \mathbf{Y}_S^T+\mathbf{Y}_S \mathbf{Y}_\nu^T
\right).
\end{aligned}
\label{eq:linear_ss}
\end{equation}

\begin{figure}
    \centering
    \includegraphics[width=0.55\linewidth]{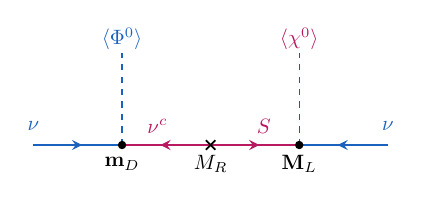}
    \caption{Diagram illustrating neutrino mass generation in the linear seesaw mechanism.}
    \label{fig:linear_ss}
\end{figure}

This structure can be understood from the mass-insertion diagram shown in Fig.~\ref{fig:linear_ss}. The same neutrino mass generation mechanism emerges, with different symmetry assumptions, in the type Ib seesaw construction proposed in~\cite{Fu:2021fyk}.  The active neutrino first couples to the singlet $\nu^c$ through the insertion $v_\Phi \mathbf{Y}_\nu$, while $\nu^c$ and $S$ are connected by the Dirac mass $M_R$ and the singlet $S$ in turn couples back to the active neutrino through the insertion $v_\chi \mathbf{Y}_S$. 
At energies below $M_R$, this chain induces the effective operator $(L\Phi)(L\chi_L)/M_R$. Notice that in our case the light-neutrino masses vanish \textit{linearly} as $v_\chi\to 0$ (hence the name linear seesaw mechanism~\cite{Akhmedov:1995ip,Akhmedov:1995vm,Malinsky:2005bi}), and lepton number is restored. Consequently, tiny active neutrino masses are  natural (symmetry-protected) and arise from the small value of the vacuum expectation value (VEV) $v_{\chi}$ of the neutral component of the leptophilic scalar doublet $\chi_L$.

\subsection{Explicit soft lepton-number breaking}

In this section we focus on explicit soft breaking of lepton number in the simplest linear seesaw scenario. The most general renormalizable scalar potential, invariant under the SM gauge symmetry and consistent with a soft breaking of the lepton number symmetry $U(1)_L$, can be written as 
\begin{equation}
\begin{aligned}
V
=&
-\mu_\Phi^2\Phi^\dagger\Phi
-\mu_\chi^2\chi_L^\dagger\chi_L
+\lambda_1(\Phi^\dagger\Phi)^2
+\lambda_2(\chi_L^\dagger\chi_L)^2
+\lambda_3(\chi_L^\dagger\chi_L)(\Phi^\dagger\Phi)
\\
&+
\lambda_4\chi_L^\dagger\Phi\Phi^\dagger\chi_L
-
\left(
\mu_{12}^2\Phi^\dagger\chi_L+\mathrm{h.c.}
\right).
\end{aligned}
\end{equation}
For simplicity, we assume a CP-conserving scalar potential, which implies that all its parameters are real.
The quartic terms conserve lepton number, whereas the bilinear term proportional to $\mu_{12}^2$ explicitly breaks $U(1)_L$ by two units in a soft way. In the limit $\mu_{12}^2\to 0$, lepton number is restored. 
This soft-breaking term plays two important roles. First, it induces a nonzero vacuum expectation value for the leptophilic doublet $\chi_L$. Second, by explicitly breaking lepton number, it avoids the appearance of a physical Majoron associated with the spontaneous breaking of a global symmetry. 
We use the convention 
\begin{equation}
\begin{aligned}
\tan\beta=\frac{v_\Phi}{v_\chi},
\qquad
v_\Phi=v\sin\beta,
\qquad
v_\chi=v\cos\beta.
\end{aligned}
\label{beta}
\end{equation}
Therefore, the neutrinophilic regime $v_\chi\ll v_\Phi$ corresponds to $\beta\simeq \pi/2$. 

After electroweak symmetry breaking, we expand the scalar doublets as
\begin{equation}
\begin{aligned}
\Phi
=
\begin{pmatrix}
\Phi^+
\\[1mm]
\dfrac{1}{\sqrt{2}}\left(v_\Phi+\Phi_R+i\Phi_I\right)
\end{pmatrix},
\qquad
\chi_L
=
\begin{pmatrix}
\chi^+
\\[1mm]
\dfrac{1}{\sqrt{2}}\left(v_\chi+\chi_R+i\chi_I\right)
\end{pmatrix}.
\end{aligned}
\end{equation}
In the CP-even sector, the physical states are defined as 
\begin{equation}
\label{eq:physical_higgs}
\begin{aligned}
h
&=
c_\alpha \Phi_R-s_\alpha \chi_R,
&
H
&=
s_\alpha \Phi_R+c_\alpha \chi_R,
\end{aligned}
\end{equation}
where $c_x\equiv\cos x$ and $s_x\equiv\sin x$. We identify $h$ with the SM Higgs boson, while $H$ corresponds to the additional CP-even scalar. In the CP-odd and charged sectors, we have 
\begin{equation}
\label{eq:physical_leptophilic}
\begin{aligned}
A
&=
c_\beta \Phi_I-s_\beta \chi_I,
&
H^-
&=
c_\beta \Phi^- -s_\beta \chi^- .
\end{aligned}
\end{equation}
The orthogonal states correspond to the Goldstone bosons absorbed by the longitudinal components of the $Z$ and $W^\pm$ gauge bosons. In the limit $v_\chi\ll v_\Phi$, the heavy scalars $H,A,H^\pm$ are dominated by the leptophilic component $\chi_L$, while $h$ remains predominantly aligned with the SM doublet. 

The masses of the physical pseudoscalar and charged scalars are given by
\begin{equation}
\begin{aligned}
m_A^2
&=
\mu_{12}^2\frac{v^2}{v_\Phi v_\chi},
\\
m_{H^\pm}^2
&=
v^2
\left(
\frac{\mu_{12}^2}{v_\Phi v_\chi}
-\frac{\lambda_4}{2}
\right).
\end{aligned}
\end{equation}
Notice that, equivalently,
\begin{equation}
\begin{aligned}
\mu_{12}^2
=
m_A^2\frac{v_\Phi v_\chi}{v^2},
\end{aligned}
\end{equation}
showing that the explicit lepton-number-breaking parameter $\mu_{12}^2$ induces a small $v_\chi$, while keeping the physical pseudoscalar mass adequately large. 

For the phenomenological analysis, it is convenient to express the quartic couplings in terms of physical masses, mixing angles and VEVs. With the conventions above, one obtains 
\begin{equation}
\label{eq:lambda_parameters}
\begin{aligned}
\lambda_1
&=
\frac{1}{2v^2\sin^2\beta}
\left(
m_H^2\sin^2\alpha
+
m_h^2\cos^2\alpha
-
m_A^2\cos^2\beta
\right),
\\
\lambda_2
&=
\frac{1}{2v^2\cos^2\beta}
\left(
m_h^2\sin^2\alpha
+
m_H^2\cos^2\alpha
-
m_A^2\sin^2\beta
\right),
\\
\lambda_3
&=
\frac{1}{v^2}
\left[
2m_{H^\pm}^2
-
m_A^2
+
\frac{(m_H^2-m_h^2)\sin2\alpha}{\sin2\beta}
\right],
\\
\lambda_4
&=
\frac{2}{v^2}
\left(
m_A^2-m_{H^\pm}^2
\right).
\end{aligned}
\end{equation}

At tree level, vacuum stability requires 
\begin{equation}
\label{eq:vacumm_stability}
\begin{aligned}
\lambda_1\geq 0,
\qquad
\lambda_2\geq 0,
\qquad
\lambda_3\geq -2\sqrt{\lambda_1\lambda_2},
\qquad
\lambda_3+\lambda_4\geq -2\sqrt{\lambda_1\lambda_2}.
\end{aligned}
\end{equation}
In addition, in the numerical analysis we impose perturbativity of the quartic couplings,
\begin{equation}
\label{eq:perturbativity}
\begin{aligned}
|\lambda_i|<\lambda_{\rm pert}=4\pi.
\end{aligned}
\end{equation}

\subsection{Neutrino masses}

As mentioned, here we focus on the minimal realisation containing only one pair of singlet mediators, as it will be the most predictive.
Following the general discussion given in~\cite{Schechter:1980gr} this provides an example of missing partner seesaw mechanism, as the matrix $\mathbf{m}_\nu$ has rank two, so that one of the active neutrinos remains massless. For example, in the case of normal ordering, the two nonzero masses can be written as  
\begin{equation}
\label{eq:rho_r}
\begin{aligned}
|m_2|
&=
\frac{v_\Phi v_\chi}{2M_R}
y_\nu y_S(1-\rho),
\qquad
|m_3|=
\frac{v_\Phi v_\chi}{2M_R}
y_\nu y_S(1+\rho),
\end{aligned}
\end{equation}
where
\begin{equation}
\label{eq:neutrino_masses}
\begin{aligned}
\rho
=
\frac{\sqrt{1+r}-\sqrt{r}}
{\sqrt{1+r}+\sqrt{r}},
\qquad
r=
\frac{|\Delta m_{21}^2|}
{|\Delta m_{32}^2|},
\end{aligned}
\end{equation}
a parameter well-determined by global neutrino oscillation fits~\cite{deSalas:2020pgw}. Eqs.~\eqref{eq:rho_r} and \eqref{eq:neutrino_masses} for inverted ordering can be found in \cite{Mandal:2026wbz}. From Eq.~\eqref{eq:neutrino_masses}, one has
\begin{equation}
\label{eq:sum_neutrino_masses}
    \Sigma m_\nu=|m_2|+|m_3|=\frac{v_\Phi v_\chi}{M_R}y_\nu y_S,
\end{equation}
which fixes the product $y_\nu y_S$ once $v_\chi$, $M_R$, and the sum of neutrino masses are specified\footnote{Combining various cosmological observations \cite{Planck:2018vyg,Naredo-Tuero:2024sgf,DESI:2025zgx,Jiang:2024viw} can lead to tight limits on the sum of the active neutrino masses, e.g. $\Sigma m_\nu \lesssim 64\text{--}77\,\mathrm{meV}$.}.\\

The observed oscillation data fix the product of Yukawa couplings $y_\nu y_S$
\begin{equation}
y_\nu y_S=\frac{M_R\sum m_\nu}{v_\Phi v_\chi},
\label{eq:yukawa-product-lepto}
\end{equation}
while the ratio $r_y$ remains as a free parameter
\begin{equation}
r_y=\frac{y_S}{y_\nu}.
\label{eq:yukawa-ratio-lepto}
\end{equation}
The Yukawa vectors determining $\mathbf{m}_D$ and $\mathbf{M}_L$ in Fig.~\ref{fig:linear_ss} and Eq.~\eqref{eq:linear_ss} can be written as\\
\begin{equation}
\mathbf Y_\nu=y_\nu\widehat{\boldsymbol n},
\qquad
\mathbf Y_S=y_S\widehat{\boldsymbol s},
\label{Yukawavectors}
\end{equation}\\
where $\widehat{\boldsymbol n}$ and $\widehat{\boldsymbol s}$ are complex unit vectors in lepton-flavour space\\
\begin{equation}
    \sum_i|\widehat n_i|^2=\sum_i|\widehat s_i|^2=1.
\end{equation}\\
For normal neutrino mass ordering, the unit vectors can be written as
\begin{equation}
\begin{aligned}
\widehat{\boldsymbol n}
&=
\frac{1}{\sqrt{2}}
\left(
\sqrt{1+\rho}\,\mathbf{K}_3
+
i\sqrt{1-\rho}\,\mathbf{K}_2
\right),
\\
\widehat{\boldsymbol s}
&=
\frac{1}{\sqrt{2}}
\left(
\sqrt{1+\rho}\,\mathbf{K}_3
-
i\sqrt{1-\rho}\,\mathbf{K}_2
\right),
\end{aligned}
\label{eq:yukawa-parametrization-linear-seesaw}
\end{equation}
where $\mathbf{K}_2$ and $\mathbf{K}_3$ denote the second and third columns of the neutrino mixing matrix $\mathbf{K}$, respectively. The relative $\pm i$ phase factors do not entail any CP-violation but ensure positivity of neutrino mass eigenvalues~\cite{Schechter:1980gr}. The parameter $\rho$ is given as
$\rho=\widehat{\boldsymbol s}^\dagger\widehat{\boldsymbol n}
=
\widehat{\boldsymbol n}^\dagger\widehat{\boldsymbol s}
$.

The unitary matrix \textbf{K} is simply the lepton mixing matrix, which can be expressed as
\begin{equation}
\label{eq:symmetric_parametrization}
\begin{aligned}
\mathbf{K}=
\omega_{23}(\theta_{23},\phi_{23})
\omega_{13}(\theta_{13},\phi_{13})
\omega_{12}(\theta_{12},\phi_{12}).
\end{aligned}
\end{equation}
In this form~\cite{Rodejohann:2011vc}, the rephasing-invariant Dirac phase is given by
\begin{equation}
\begin{aligned}
\delta_{\rm CP}
=
\phi_{13}-\phi_{12}-\phi_{23}.
\end{aligned}
\end{equation}  
Note that, in the symmetrical parametrization adopted here, the CP-violating structure relevant for both low-energy observables and leptogenesis in the minimal linear seesaw is controlled by the Dirac phase $\delta_{\rm CP}$ and a single physical Majorana phase.
The unitary matrix $\mathcal{U}$ diagonalizing the full $5\times 5$ neutral lepton mass matrix of Eq.~\eqref{eq:full_neutrino_matrix}, $\mathcal{U}^\dagger\mathcal{M}_\nu \mathcal{U}^\ast=\mathcal{M}_\nu^{\rm diag}$, can be written as 
\begin{equation}
\begin{aligned}
\mathcal U
\simeq
\begin{pmatrix}
\left(1-\dfrac{1}{2}\mathbf S\mathbf S^\dagger\right)\mathbf K
&
-\dfrac{i}{\sqrt{2}}\mathbf S
&
\dfrac{1}{\sqrt{2}}\mathbf S
\\[2mm]
0
&
\dfrac{i}{\sqrt{2}}\left(1-\dfrac{1}{2}\mathbf S^\dagger\mathbf S\right)
&
\dfrac{1}{\sqrt{2}}\left(1-\dfrac{1}{2}\mathbf S^\dagger\mathbf S\right)
\\[2mm]
-\mathbf S^\dagger
&
-\dfrac{i}{\sqrt{2}}\left(1-\dfrac{1}{2}\mathbf S^\dagger\mathbf S\right)
&
\dfrac{1}{\sqrt{2}}\left(1-\dfrac{1}{2}\mathbf S^\dagger\mathbf S\right)
\end{pmatrix}
+
\mathcal O(\mathbf S^3).
\end{aligned}
\label{eq:full-mixing-nonunitary}
\end{equation}
At this level, the effective light block of the full mixing matrix~\cite{Schechter:1980gr} 
\begin{equation}
\label{eq:non_unitary_KU}
\begin{aligned}
\widetilde{\mathbf K}
=
\left(1-\frac{1}{2}\mathbf S\mathbf S^\dagger\right)\mathbf K,
\end{aligned}
\end{equation}
is not strictly unitary. The mixing between light and heavy fermions is given by the block
\begin{equation}
\begin{aligned}
\widetilde{\mathbf K}_H
=
\left(
-\frac{i}{\sqrt{2}}\mathbf S,\,
\frac{1}{\sqrt{2}}\mathbf S
\right).
\end{aligned}
\end{equation}

The unitarity of $\mathcal{U}$ implies
\begin{equation}
\begin{aligned}
\widetilde{\mathbf K}\widetilde{\mathbf K}^\dagger
+
\widetilde{\mathbf K}_H\widetilde{\mathbf K}_H^\dagger
=
1
+
\mathcal O(\mathbf S^4).
\end{aligned}
\end{equation}
At leading order, the mixing matrix is reduced to 
\begin{equation}
\label{eq:neutrino_mixing_small}
\begin{aligned}
\mathcal{U}
\simeq
\begin{pmatrix}
\mathbf{K} & -\dfrac{i}{\sqrt{2}}\mathbf{S} & \dfrac{1}{\sqrt{2}}\mathbf{S}
\\[2mm]
0 & \dfrac{i}{\sqrt{2}} & \dfrac{1}{\sqrt{2}}
\\[2mm]
-\mathbf{S}^\dagger & -\dfrac{i}{\sqrt{2}} & \dfrac{1}{\sqrt{2}}
\end{pmatrix}+
\mathcal O(\mathbf S^2).
\end{aligned}
\end{equation}

Both the doublet-singlet neutrino mixing and non-unitarity are controlled by the dimensionless quantity 
\begin{equation}
\begin{aligned}
\mathbf{S}
\simeq
\frac{\mathbf{m}_D+\mathbf{M}_L}{M_R}.
\end{aligned}
\end{equation}
In the low-$v_\chi$ regime, where $|\mathbf{M}_L|\ll |\mathbf{m}_D|$, the ratio $v_\chi y_S/(v_\Phi y_\nu)$ is suppressed, and $\mathbf{S}$ reduces to the small seesaw expansion parameter of Ref.~\cite{Schechter:1981cv},
\begin{equation}
\label{eq:S_mixing_matrix}
\begin{aligned}
\mathbf{S}
\simeq
\frac{\mathbf{m}_D}{M_R}
=
\frac{v_\Phi}{\sqrt{2}M_R}\,y_\nu\widehat{\boldsymbol n} .
\end{aligned}
\end{equation}
Therefore, the mixing of the singlet mediators with the active states is suppressed by the ratio $|\mathbf{m}_D|/M_R$. In the decoupling limit $M_R\to\infty$, with Yukawa couplings fixed, one has $\mathbf{S}\to 0$.   

The two heavy neutral mediator fermions form a quasi-Dirac pair, denoted by $N_4$ and $N_5$. Their masses are approximately
\begin{equation}
\begin{aligned}
M_{N_{4,5}}
\simeq
M_R
\left(
1+\frac{1}{2}|\mathbf{S}|^2
\right)
\mp
\frac{1}{2M_R}
\left(
\mathbf{M}_L^\dagger\mathbf{m}_D
+
\mathbf{m}_D^\dagger\mathbf{M}_L
\right).
\end{aligned}
\label{eq:heavy-neutrino-masses}
\end{equation} 
Therefore, the mass splitting of the quasi-Dirac pair is controlled by the same lepton-number-violating term responsible for the light-neutrino masses,
\begin{equation}
\label{eq:mass_splitting}
\begin{aligned}
\Delta M_N
=
\frac{v_\Phi v_\chi}{M_R}
\left|\mathbf{Y}_S^\dagger\mathbf{Y}_\nu
\right|=\frac{v_\Phi v_\chi}{M_R}y_\nu y_S\rho.
\end{aligned}
\end{equation}

In the lepton-number-conserving limit, $v_\chi\to 0$, neutrinos become massless and the mass splitting between the mediators $N_4$ and $N_5$ vanishes, so they combine into an exact Dirac fermion. Consequently, the minimal linear seesaw naturally provides a pair of quasi-degenerate heavy neutrinos. This opens the way for a \textit{testable} theory of low-scale resonant leptogenesis, obtained when the splitting $\Delta M_N$ is comparable to the decay widths of $N_4$ and $N_5$. \\ 

\section{Leptogenesis}
\label{sec:leptogenesis}

The minimal linear seesaw model contains a single quasi-Dirac pair of mediators, $N_4$ and $N_5$, whose mass splitting is fixed at leading order by the light-neutrino spectrum,
$$
\Delta M_N=\rho\,\Sigma m_\nu,$$ see Eqs.~\eqref{eq:sum_neutrino_masses}
and \eqref{eq:mass_splitting}. 
This provides a natural theory setup for resonant leptogenesis. The resonant enhancement becomes sizeable when the heavy-neutrino decay width is comparable to the mass splitting, with maximal enhancement at $\Gamma\simeq2\Delta M_N$.

\begin{figure}[b]
    \centering
    \subfigure[]{
        \centering
        \includegraphics[width=0.3\linewidth]{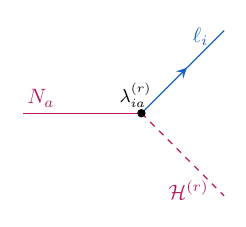}
    }
    \subfigure[]{
        \centering
        \includegraphics[width=0.3\linewidth]{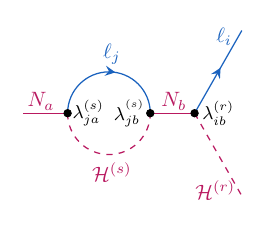}
    }
    \subfigure[]{
        \centering
        \includegraphics[width=0.3\linewidth]{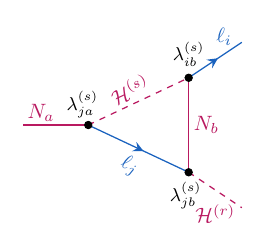}
    }
    \caption{Tree-level, self-energy and vertex contributions to the heavy-neutrino decay asymmetries. Here the indices $r,s=h,H,A,H^\pm$ label the different scalar channels and $\ell$ may correspond to a charged lepton or a neutrino.}
    \label{fig:cp_decays}
\end{figure}

The out-of-equilibrium heavy-neutrino decay Feynman diagrams are the conventional ones, shown in Fig.~\ref{fig:cp_decays}.
In a leptogenesis scenario, the weak washout regime strongly depends on the initial plasma conditions, and approximate treatments of it are generally unreliable, particularly in the resonant regime, where a careful thermal analysis is required. In contrast, the strong washout regime makes the predicted baryon asymmetry largely insensitive to initial conditions and thus allows reliable predictions. 

In this work, we focus on the strong washout regime. 
In this regime, decays and inverse decays drive the abundances of the quasi-degenerate states $N_4$ and $N_5$ close to their equilibrium values by the epoch $T\sim M_R$, making the final result largely insensitive to their initial abundances. 
As the Universe cools below this temperature, the equilibrium heavy-neutrino abundances become Boltzmann suppressed and their CP-asymmetric decays source the flavoured charges $Y_{\Delta_i}\equiv Y_{B/3-L_i}$. As we see below, the CP source is intrinsically flavoured, implying that the non-universal inverse-decay washout is necessary to obtain a non-vanishing final asymmetry compatible with cosmological observations. 

In the strong washout regime, the late-time asymmetry is accumulated mainly around the inverse-decay freeze-out epoch $z_B\equiv M_R/T_B$, when the inverse-decay washout becomes Boltzmann suppressed with respect to the Hubble expansion and the heavy-neutrino population is depleted. While sphalerons remain in equilibrium, they continuously reprocess the flavoured $B-L$ charges into baryon number; once their rate becomes slower than the expansion rate near $T_{\rm sph}$, the baryon asymmetry freezes out~\cite{Eijima:2017anv}. The objective of this work is therefore to determine whether the surviving baryon asymmetry after sphaleron freeze-out can reproduce the value indicated by the Planck Collaboration \cite{Planck:2018vyg}
\begin{equation}
\label{eq:obs_YB}
Y_{\Delta B}^{\rm obs}
=(8.69\pm0.04)\times10^{-11}.
\end{equation}

The decay channels involved in determining the baryon asymmetry are
\begin{equation}
N_a \to \nu_i h,\ \nu_i H,\ \nu_i A,
\qquad
N_a \to \ell_i^\mp H^\pm,
\end{equation}
where $a=4,5$ labels the heavy mediators and $i=e,\mu,\tau$ denotes the lepton flavour. The CP asymmetry arises from the interference between the tree-level amplitude and the absorptive part of the one-loop corrections in Figure~\ref{fig:cp_decays}. In this model the CP-source can be written in the form 
\begin{equation}
\epsilon_{ai}^{(r)}
\simeq
\varepsilon_b^{\mathrm{res}}\,\kappa_a^{(r)}\,\eta_i^{\mathrm{CP}},
\label{ep}
\end{equation}
where $(a,b)=(4,5),(5,4)$. For details see Appendix \ref{app:decay_cp}. This factorized form holds in our linear seesaw scheme and follows from Eqs.~\eqref{eq:resonant_cp_asymmetry} and \eqref{eq:factorization}, separating three physical ingredients: the resonant factor (also known the regulator \cite{Deppisch:2010fr, BhupalDev:2014pfm}), the scalar coefficient and the flavour structure. The coefficient $\kappa_a^{(r)}$ contains the scalar mixing, Yukawa and kinematic dependence. 

The factor $\eta_i^{\rm CP}$ is a flavoured CP-invariant quantity
\begin{equation}
\label{eq:cp_invariant}
\begin{aligned}
    \eta_i^{\rm CP}
    &\equiv
    \operatorname{Im}
    \left(
    Y_{S i}^{\ast}Y_{\nu i}
    \right)
    =
    y_\nu y_S
    \operatorname{Im}
    \left(
    \widehat {s_i}^\ast\widehat n_i
    \right).
\end{aligned}
\end{equation}

Note that the asymmetry is intrinsically flavoured. Although the individual $\eta_i^{\rm CP}$ can be nonzero, their sum vanishes:
\begin{equation}
\sum_{i=e,\mu,\tau}
\eta_i^{\rm CP}
=
y_\nu y_S
\operatorname{Im}
\left(
\widehat{\boldsymbol s}^{\dagger}
\widehat{\boldsymbol n}
\right)
=
0.
\label{eq:flavoured-cp-cancellation}
\end{equation}
The unflavoured source therefore cancels, implying that a nonzero final asymmetry requires flavour-dependent washout factors. 

This fact holds beyond leading order in light neutrino mixing. Indeed, even if one replaces the unitary matrix $\mathbf K$ by the non-unitary block defined in Eq.~\eqref{eq:non_unitary_KU}, the sum of the CP invariants becomes
\begin{equation}
\begin{aligned}
\sum_i \widetilde{\eta}_i^{\rm CP}
&=
y_\nu y_S\sqrt{1-\rho^2}\,
\operatorname{Re}
\left[
(\widetilde{\mathbf K}^\dagger\widetilde{\mathbf K})_{23}
\right]\\
&=y_\nu y_S\sqrt{1-\rho^2}\,
\operatorname{Re}
\left[
-
(\mathbf K^\dagger \mathbf S^\dagger \mathbf S\mathbf K)_{23}
\right]\\
&=0.
\end{aligned}
\end{equation}
Therefore, in the seesaw regime of the minimal model, the non-unitarity of the leptonic mixing matrix induced by doublet-singlet neutrino mixing does not generate an additional CP source for resonant leptogenesis. 

The first factor in Eq.~\eqref{ep}, which accounts for resonant enhancement, is given by
\begin{equation}
\label{eq:eres}
\varepsilon_a^{\mathrm{res}}
=
\frac{R_{\Gamma,a}}{1+R_{\Gamma,a}^2},
\qquad
R_{\Gamma,a}
=
\frac{\Gamma_a}{2\Delta M_N}.
\end{equation}
With our conventions, $\varepsilon_a^{\mathrm{res}}$ reaches its maximum value of $1/2$ when $\Gamma_a=2\Delta M_N$. The ratio $R_{\Gamma,a}$ compares the decay width with the mass splitting of the quasi-Dirac pair, allowing the resonant region to be identified with $R_\Gamma=1$.

To estimate the baryon asymmetry generated in each lepton flavour, we use the approximate solution of the Boltzmann equation, expressed as the product of the CP asymmetry in heavy-neutrino decays, $\epsilon_{ai}$, and an efficiency factor, $\eta_{ai}$, which represents the fraction of that CP asymmetry that survives washout \cite{Davidson:2008bu}. A generalization including the decays of different heavy neutrinos with different scalar channels is given by \cite{Fu:2021fyk} 
\begin{equation}
\label{eq:baryon_asymmetry}
Y_{\Delta B_i}
\simeq 
\sum_{a=4,5}
\epsilon_{ai}^{\mathrm{eff}}\eta_{ai}Y^{\rm eq}_{N_a}(z\ll 1).
\end{equation}
Here $\epsilon_{ai}^{\mathrm{eff}}$ denotes the effective CP source for each heavy neutrino and lepton flavour,
\begin{equation}
\epsilon_{ai}^{\mathrm{eff}}
=
\sum_{r=h,H,A,H^\pm}\epsilon_{ai}^{(r)},
\end{equation}
while $Y^{\rm eq}_{N_a}(z\ll 1)$ is the relativistic equilibrium abundance of a Majorana state normalised to the entropy density, which is approximately constant for temperatures much larger than the heavy neutrino mass scale $T \gg M_N$, because both $n_N^{\mathrm{eq}}$ and $s$ scale as $T^3$, and which provides the standard thermal initial condition for the Boltzmann equations \cite{Buchmuller:2004nz,Davidson:2008bu}. 
 
The washout strength is characterized by the decay parameter
\begin{equation}
\label{eq:decay_parameter}
K_{ai}
=
\frac{\Gamma_{ai}}{H(M_a)},
\qquad
\Gamma_{ai}
=
\sum_{r=h,H,A,H^\pm}\Gamma_{ai}^{(r)},
\end{equation}
where $H(M_a)$ is the Hubble expansion rate evaluated at $T=M_a$ and $\Gamma_{ai}^{(r)}$ is the decay width of the  heavy neutrino $N_a$ into the lepton $\ell_i$ and the scalar $\mathcal{H}^{(r)}$. 

The width that controls the resonance also determines the total washout. Summing the decay parameters over the final lepton flavours $K_{\mathrm{tot},a}=\sum_iK_{ai}$ we obtain from Eq.~\eqref{eq:eres}
\begin{equation}
R_{\Gamma,a}
=\frac{H(M_a)}{2\rho\sum m_\nu}\,K_{\mathrm{tot},a}.
\label{eq:resonance-washout}
\end{equation}

Thus, for fixed light and heavy neutrino spectra, the resonance selects a total washout scale. 
In the strong-washout regime, decays and inverse decays keep the heavy-neutrino abundances close to equilibrium early in their evolution. 
The surviving asymmetry builds up mainly when washout becomes inefficient. For $K_{ai}\gg1$, the efficiency of the $N_a$ heavy neutrino decaying into the lepton flavour $i$ can be approximated, following \cite{Davidson:2008bu}, as 
\begin{equation}
\eta_{ai}
\simeq
\left(
\frac{1}{2|A_{ii}|K_{ai}}
\right)^{1.16},
\label{eta}
\end{equation}
where the coefficients $A_{ii}$ account for the spectator processes \cite{Fu:2021fyk}\footnote{We use the ``$A$''$-$ matrix computed explicitly for the minimal type Ib seesaw in Eq.~(B22) of \cite{Fu:2021fyk}.}.

\begin{table*}[!b]
  \centering
  \small
  \setlength{\tabcolsep}{3pt}
  \begin{tabular}{@{}lcc@{}}
    \hline\hline
    Parameter &  & Sampling \\ \hline
    $\alpha_M/\pi$ & $[0,2)$ & Linear \\ \hline
    $v_\chi\ [\mathrm{GeV}]$ & $[10^{-4},10]$ & Logarithmic \\
    $M_R\ [\mathrm{GeV}]$ & $[150,10^4]$ & Logarithmic \\
    $m_A\ [\mathrm{GeV}]$ & $[100,10^4]$ & Logarithmic \\
    $\lambda_2$ & $[0,4\pi]$ & Linear \\
    $\lambda_3,\,\lambda_4$ & $[-4\pi,4\pi]$ & Linear \\
    $r_y=y_S/y_\nu$ & $[10^{-6},10^8]$ & Logarithmic \\
    \hline\hline
  \end{tabular}
  \caption{Parameter ranges used in our numerical scan. The oscillation parameters were fixed at the best-fit values reported in~\cite{deSalas:2020pgw} assuming symmetric uncertainties.}
  \label{tab:mls_scan_global}
\end{table*}

In order to identify the viable parameter regions that successfully reproduce the observed baryon asymmetry, we scan the relevant input parameters consistent with the neutrino oscillation data, using Eq.~\eqref{eq:neutrino_masses}.
We performed this analysis for both normal and inverted neutrino mass orderings. 
The ranges are specified in Table~\ref{tab:mls_scan_global}.
\begin{table}[!b]
\centering
\small
\begin{tabular}{lccc}
\hline
 & $\mathrm{BP}_1$ & $\mathrm{BP}_2$ & $\mathrm{BP}_3$\\
\hline
$Y_{\Delta B}$
& $8.40321\times10^{-11}$
& $8.67314\times10^{-11}$
& $8.65316\times10^{-11}$\\
$K_{\min}$
& $125.04$
& $32.5803$
& $55.3704$\\
$\{K_{\mathrm{tot},4},K_{\mathrm{tot},5}\}$
& $\{1482.99,1499.22\}$
& $\{468.945,471.531\}$
& $\{537.645,540.349\}$\\
$\{R_{\Gamma,4},R_{\Gamma,5}\}$
& $\{18.5452,18.7482\}$
& $\{12.2908,12.3586\}$
& $\{0.476639,0.479036\}$\\
\hline
$y_\nu$
& $1.39302\times10^{-5}$
& $5.93213\times10^{-6}$
& $6.16908\times10^{-6}$\\
$y_S$
& $1.11244\times10^{-5}$
& $1.86624\times10^{-5}$
& $0.0846406$\\
$v_\chi\,[\mathrm{GeV}]$
& $1.29024$
& $2.63281$
& $1.02331\times10^{-4}$\\
$M_R\,[\mathrm{GeV}]$
& $834.114$
& $1209.91$
& $223.802$\\
$m_H\,[\mathrm{GeV}]$
& $2930.43$
& $1124.37$
& $520.33$\\
$m_A\,[\mathrm{GeV}]$
& $2930.43$
& $1124.42$
& $520.33$\\
$m_{H^\pm}\,[\mathrm{GeV}]$
& $2883.36$
& $991.973$
& $635.694$\\
\hline
\end{tabular}
\caption{Three reference benchmark values with normal neutrino mass ordering. The pairs of values of $K_{\mathrm{tot}}$ and $R_\Gamma$ correspond to the two heavy neutrino mediators.}
\label{tab:leptogenesis-benchmarks}
\end{table}
We fix $v=246.22\,\mathrm{GeV}$ and $m_h=125.13\,\mathrm{GeV}$ \cite{ParticleDataGroup:2026mpi}, and reconstruct $v_\Phi$ from Eq.~\eqref{eq:scalar_doublets}, the scalar masses and mixing angles as well as the scalar potential parameters from Eq.~\eqref{eq:lambda_parameters}.
We impose perturbativity on the quartic scalar couplings and the tree-level vacuum stability conditions from Eq.~\eqref{eq:vacumm_stability}. We require that the resulting diphoton Higgs signal strength values $R_{\gamma\gamma}$, defined in Ref. \cite{Mandal:2026wbz}, lie within their $3\sigma$ experimentally allowed range \cite{ATLAS:2026pdi,CMS:2026zxu}.
In the leptonic sector we require $\max_i\{|Y_{\nu i}|,|Y_{Si}|\}<\sqrt{4\pi}$. The selected points satisfy $Y_{\Delta B}>0$ and $K_{\min}=\min_{a,i}K_{ai}\gg1$ so that all three flavours are in the strong-washout regime for both heavy neutrinos. 

We compute $Y_{\Delta B}$ using the analytical approximation in Eq.~\eqref{eq:baryon_asymmetry}, summing the contributions from $N_4$ and $N_5$ over the lepton flavours $e,\mu,\tau$ and the kinematically allowed scalar channels $h,H,A,H^\pm$. We use $c_{\rm sph}=10/31$ and $Y_N^{\rm eq}(0)=45/(\pi^4g_*)$, with $g_*=114.25$. We identify successful points compatible with the observed baryon asymmetry of the Universe in Eq.~\eqref{eq:obs_YB} by requiring $0.9\leq Y_{\Delta B}/Y_{\Delta B}^{\rm obs}\leq1.1$. 

\subsection{Numerical results}

The model's simplicity is reflected in the relatively simple dependence of asymmetry formulae (see Eqs.~\eqref{eq:aprox_asym_h} and \eqref{eq:asymmetry_all_channels} below) on the Yukawa hierarchy, the scalar mass spectrum, and the neutrino oscillation data. 
The interplay among these allows us to identify different parameter-space regimes that successfully reproduce the observed baryon asymmetry of the Universe. 
To illustrate these different regimes, we choose the benchmark points BP${}_1$ and BP${}_2$ in Table~\ref{tab:leptogenesis-benchmarks}, which serve as illustrative examples for the scenario with $v_\chi\sim \mathcal{O}(1)$ GeV. In BP${}_1$, CP violation is generated solely by the decay into the SM Higgs channel, whereas in BP${}_2$ all scalar channels are kinematically open. The benchmark point BP${}_3$ was selected as a realisation of resonant leptogenesis in the $v_\chi\ll 1$ GeV regime for NO.

\begin{figure}[!b]
    \centering
    \subfigure[]{
        \centering
        \includegraphics[width=0.6\linewidth]{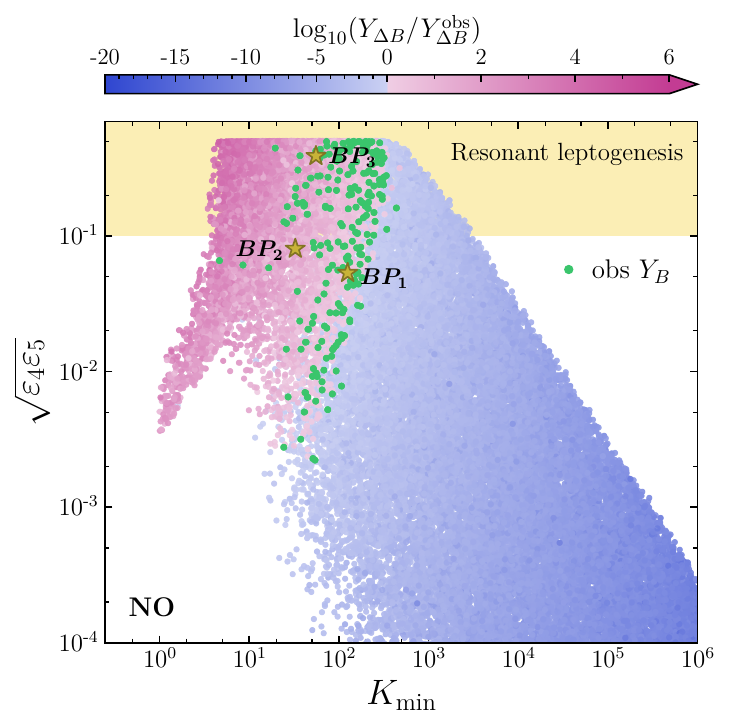}
    }  
    \subfigure[]{
        \centering
        \includegraphics[width=0.61\linewidth]{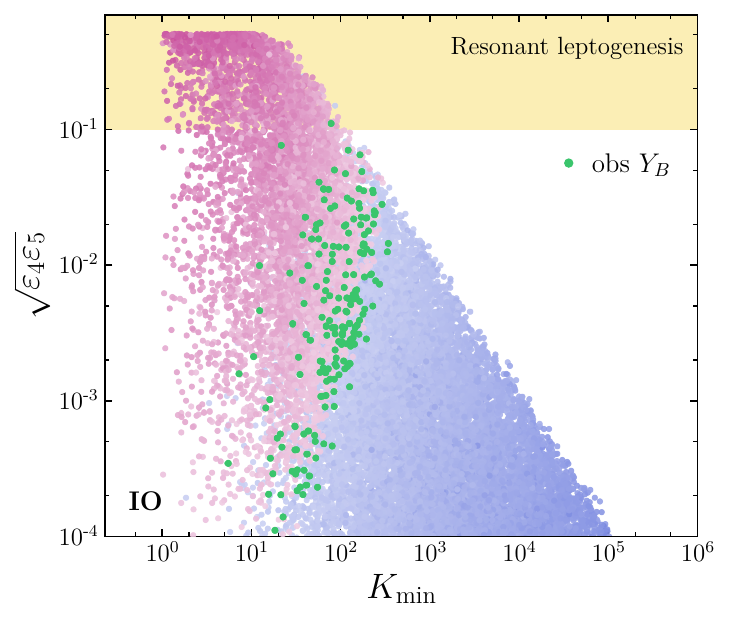}
    }
    \caption{Cosmological baryon asymmetry versus the smallest of the decay parameters $K_{\min}$ and $\varepsilon^{\rm res}$ (see Eq.~\eqref{eq:eres}) plane for NO and IO obtained in the global scan. Colours indicate $Y_{\Delta B}/Y_{\Delta B}^{\rm obs}$, green points lie in the observed range, and stars mark the benchmarks. Here BP${}_1$ and BP${}_3$ lie in the Higgs regime, while BP${}_2$ is in the leptophilic regime (see below).}
    \label{fig:K_res}
\end{figure}

Fig.~\ref{fig:K_res} shows the scan in the 
$(K_{\min},\sqrt{\varepsilon_4^{\mathrm{res}}\varepsilon_5^{\mathrm{res}}})$ plane, which allows us to study both the proximity to resonance and the washout strength. For normal neutrino mass ordering, we find solutions that reproduce the observed asymmetry both near the resonant maximum and outside that region. Among these, $\mathrm{BP}_3$ shows that resonant leptogenesis can occur with a small value of $v_\chi$ and heavy neutrinos accessible to collider energies.

A comparison with inverted ordering reveals an absence of successful resonant solutions in the explored region, although solutions are found outside of it.  This difference has a parametric interpretation, which is explained in the following. In the minimal model,
\begin{equation}
\Delta M_N^{\mathrm{NO}}=m_3-m_2,
\qquad
\Delta M_N^{\mathrm{IO}}=m_2-m_1.
\end{equation}
The smaller separation in inverted ordering implies that, for fixed $M_R$, the resonance factor selects a small total washout in comparison with NO, according to Eq.~\eqref{eq:resonance-washout}. This restricts the coexistence of resonance and strong washout, although the final asymmetry also depends on the source and its distribution among flavours. 

The comparison shows that resonance alone does not determine the result. The CP source, with its complete couplings and the efficiencies of the three flavours, also plays a role. Similarly, $K_{\min}$ identifies only the smallest decay parameter and does not capture the entire washout distribution. Solutions with very large values of $K_{\min}$ predominantly have small asymmetries, and are indicated in the blue region of Fig.~\ref{fig:K_res}, where the efficiency factor of Eq.~(\ref{eta}) is very small. 

To obtain an analytical understanding of the results of our global scan, we simplify the relevant expressions by taking $M_4\simeq M_5\simeq M_R$ and assume small scalar mixing \footnote{The mass splitting $\Delta M_N$ is nevertheless retained both in the resonant enhancement factor and in the leading scalar-mixing contribution to the CP-violating term.}.
Under these assumptions, the two heavy neutrinos have approximately equal decay widths per channel, $\Gamma_{4i}^{(r)} 
\simeq
\Gamma_{5i}^{(r)}
\equiv
\Gamma_i^{(r)}$, in consequence, 
 the CP-asymmetries and efficiencies are also equal $\epsilon_{4i}^{(r)}
\simeq
\epsilon_{5i}^{(r)}
\equiv
\epsilon_i^{(r)}$, $\eta_{4i}
\simeq
\eta_{5i}
\equiv
\eta_i$ as can be seen from Eqs.~\eqref{eq:eres} and \eqref{eq:decay_parameter}.

In the approximation described above, the factor $\kappa^{(r)}$ takes a simplified form
\begin{equation}
\kappa^{(r)}\simeq
-\frac{1}{y_\nu y_S}
\frac{P_h/r_y-r_y(P_H+P_A+2P_{H^\pm})}
{[P_h/r_y+r_y(P_H+P_A+2P_{H^\pm})]^2}
\begin{cases}
P_h\sin2\alpha, & r=h,\\
-P_H\sin2\alpha, & r=H,\\
P_A\sin2\beta, & r=A,\\
2P_{H^\pm}\sin2\beta, & r=H^\pm
\end{cases}
\label{eq:kappa-multichannel}
\end{equation}
where 
\begin{equation}
P_{r}=\left(1-\frac{m_r^2}{M_R^2}\right)^2\Theta(M_R-m_r)
\end{equation}
is the kinematic factor of the heavy neutrino decaying into the scalar $\mathcal{H}^{(r)}$. More details are given in Appendix \ref{app:decay_cp}.

\subsection{Leptogenesis in the Higgs regime}

In this subsection we analyze the regime where leptogenesis arises mainly from the heavy-neutrino decays into the SM Higgs boson, i.e. we assume the condition $m_h<M_R<\min(m_H,m_A,m_{H^\pm})$.  This implies that the CP asymmetry is sourced simply by the $N_a\to\nu_i h$ decay. 
This results in $P_H=P_A=P_{H^\pm}=0$, so using Eqs.~\eqref{eq:eres},
\eqref{eq:decay_parameter} and \eqref{eq:kappa-multichannel}
one gets the following relations
\begin{equation}
\begin{aligned}
R_\Gamma&\simeq
\frac{M_R^2P_h}{64\pi\rho v_\Phi v_\chi r_y},\quad 
K_i&\simeq
\frac{M_{\mathrm{Pl}}\sum m_\nu}{32\pi v_\Phi v_\chi}
\sqrt{\frac{45}{4\pi^3g_*}}\,
\frac{P_h}{r_y}|\hat n_i|^2,\quad
\kappa^{(h)}\simeq-\frac{r_y\sin2\alpha}{y_\nu y_S}.
\end{aligned}
\label{eq:h-ingredients}
\end{equation}

Substituting this into the baryon asymmetry, we obtain 
\begin{equation}
\colorbox{yellow!50}{$\displaystyle
Y_{\Delta B_i}
\propto
\frac{R_\Gamma}{1+R_\Gamma^2}\,
\sin2\alpha\, r_y^{2.16}\,
\frac{\operatorname{Im}(\hat s_i^*\hat n_i)}
{|A_{ii}|^{1.16}|\hat n_i|^{2.32}}.$}
\label{eq:aprox_asym_h}
\end{equation}

The expression for the baryon asymmetry involves the resonance parameter $ R_\Gamma$, the scalar mixing angle $\alpha$, the Yukawa ratio $r_y$ (given in Eq.~\eqref{eq:yukawa-ratio-lepto}) as well as a flavour-dependent parameter, coming from the complex Yukawa unit vectors $\hat s_i$ and $\hat n_i$ ($i=e,\mu,\tau$) in Eqs.~\eqref{Yukawavectors} and \eqref{eq:yukawa-parametrization-linear-seesaw}. Eq.~\eqref{eq:h-ingredients}  shows that increasing $r_y$ reduces both $ K_i$ and the resonance parameter $ R_\Gamma$ accordingly.
To illustrate this scenario, we consider BP${}_1$ from Table~\ref{tab:leptogenesis-benchmarks}. Here $r_y\simeq0.799<1$ and $v_\chi\simeq1.29\,\mathrm{GeV}$.

This benchmark illustrates how the observed baryon asymmetry can be reproduced through a decay channel predominantly aligned with the SM doublet, without requiring maximal resonance. 
Here we find $R_\Gamma\simeq 18.6$, which corresponds to a factor $\varepsilon_a^{\mathrm{res}}\simeq 5.36\times 10^{-2} \ll \frac{1}{2}$, implying that the point is far from the resonant regime. 

The CP asymmetry in this model is intrinsically flavoured. For the BP${}_1$ benchmark the decay parameters and efficiencies satisfy the following pattern
\begin{equation}
K_{ae}<K_{a\mu}<K_{a\tau},
\qquad
\eta_{ae}>\eta_{a\mu}>\eta_{a\tau}.
\label{HierarchyKandeta}
\end{equation}
The differences between $N_4$ and $N_5$ are small compared with those among flavours. In particular, the electron flavour has the weakest washout and retains a larger fraction of the produced asymmetry. 
We note that, even if a given flavour contributes only a small fraction of the CP asymmetry, differences in sign and in efficiency can nevertheless make that flavour’s final baryonic asymmetry dominant. 

\subsection{ Leptogenesis in the leptophilic regime}

We now turn to the leptophilic regime in which the heavy-neutrino mediators decay mainly to the leptophilic scalars. For $M_R>\max(m_h,m_H,m_A,m_{H^\pm})$, the additional scalars contribute to both the CP-factor and the decay parameter. Here we discuss analytically the results of our global parameter scan associated with the benchmark point BP${}_2$ which reproduces the observed baryon asymmetry when all the scalar channels are open.
In this scenario, the decay rate, the washout and the resonant parameters are  given by 
\begin{equation}
\begin{aligned}
&
\Gamma
\simeq
\frac{M_R^2\Sigma m_\nu}
{32\pi v_\Phi v_\chi}
\left[
\frac{P_h}{r_y}
+
r_y
\left(
P_H+P_A+P_{H^\pm}
\right)
\right],\\
&
K_i
\simeq
\frac{M_{\rm Pl}\Sigma m_\nu}
{32\pi v_\Phi v_\chi}
\sqrt{\frac{45}{4\pi^3g_\ast}}
\left[
\frac{P_h|\widehat n_i|^2}{r_y}
+
r_y
\left(
P_H+P_A+P_{H^\pm}
\right)
|\widehat s_i|^2
\right]
,\\
&
R_\Gamma
\simeq
\frac{M_R^2}
{64\pi\rho\,v_\Phi v_\chi}
\left[
\frac{P_h}{r_y}
+
r_y
\left(
P_H+P_A+P_{H^\pm}
\right)
\right]
\label{eq:chi_ingredients}
\end{aligned}
\end{equation}
respectively. 
The baryon asymmetry can be written as
\begin{equation}
\colorbox{yellow!50}{$\displaystyle
\begin{aligned}
Y_{\Delta B_i}
\propto\,&
\frac{R_\Gamma}{1+R_\Gamma^2}
\frac{P_h/r_y-r_y(P_H+P_A+P_{H^\pm})}
{\left[P_h/r_y+r_y(P_H+P_A+P_{H^\pm})\right]^2}\left[(P_h-P_H)\sin2\alpha
+(P_A+P_{H^\pm})\sin2\beta\right]\\
&\times\frac{
\operatorname{Im}(\hat s_i^*\hat n_i)
}{
|A_{ii}|^{1.16}
\left[
P_h|\hat n_i|^2/r_y
+
r_y(P_H+P_A+2P_{H^\pm})|\hat s_i|^2
\right]^{1.16}
}.
\end{aligned}$}
\label{eq:asymmetry_all_channels}
\end{equation}

Now the asymmetry, the resonance, and the decay parameter are controlled by the competition among the four kinematical factors $P_h$, $P_H$, $P_A$, $P_{H^\pm}$ and the mixing angle $\alpha$ (note that $\beta \simeq \pi/2$, a consequence of Eq.~(\ref{beta}) and the hierarchy $v_\chi \ll v_\Phi$). When the leptophilic channels dominate, raising $r_y$ enhances the decay width $\Gamma$ as well as the decay and resonant parameters $K_i$ and $R_{\Gamma}$, rather than suppressing them. This follows from Eq.~\eqref{eq:chi_ingredients}, which shows that the contributions arising from heavy non-SM scalar states to the decay rate, the washout, and the resonant parameters depend linearly on the Yukawa coupling ratio $r_y$.

As an example, we use the benchmark point BP${}_2$. Here $r_y\simeq 3.15>1$ and $v_\chi\simeq 2.63$ GeV. The mass spectrum favours the charged channel over the new neutral ones. The kinematical factors are $P_H\simeq P_A\simeq0.0186$, implying that the $H^\pm$ channel provides the dominant contribution to the baryon asymmetry of the Universe.

For this benchmark, we obtain
\begin{equation}
K_{ae}<K_{a\tau}<K_{a\mu},
\qquad
\eta_{ae}>\eta_{a\tau}>\eta_{a\mu}.
\end {equation}
The electron flavour retains the highest efficiency, while the relative order of the muonic and tauonic flavours is reversed compared to the previous scenario. 
Eq.~\eqref{eq:chi_ingredients} shows how $|\hat s_i|^2$ can modify the order of the flavour contributions. The result depends on the flavour pattern. Once again, different efficiencies play an important role in determining the final asymmetry.
The benchmark BP${}_{3}$ corresponds to a successful point in the scan for light neutrinos with normal mass ordering in the region of small $v_\chi$ and a pronounced Yukawa hierarchy \bl{$r_y$}.

In short, we found that acceptable values of the observed cosmological baryon asymmetry can be obtained within the resonant leptogenesis scenario. 
Successful solutions can be found in the regime where the heavy neutrinos decay predominantly into the leptophilic scalar rather than the SM Higgs boson.

\subsection{Phenomenology}

The first interesting phenomenological feature of our Linear Seesaw leptogenesis framework concerns the lepton number violating neutrinoless double-beta decay process (\znbb for short), a characteristic consequence of having \lnv and neutrinos as Majorana fermions~\cite{Schechter:1981bd}.
The parametrization of Eq.~\eqref{eq:symmetric_parametrization} provides a neat description of the \znbb decay amplitude, 
whose relevant effective mass parameter is given as~\cite{Rodejohann:2011vc}  
\begin{equation}
\label{eq:mbb}
|m_{\beta\beta}|  = \left|\sum_{j=1}^3 U_{ej}^2 m_j\right|=\left|c^{2}_{12}c^{2}_{13} m_1 + s^{2}
_{12}c^{2}_{13} m_2 e^{-2i\phi_{12} }+ s^{2}_{13} m_3 e^{-2i\phi_{13}}\right|~.
\end{equation}
There are important restrictions on the attainable values of $|m_{\beta\beta}|$ that follow from the existing oscillation experiments, as well as direct \znbb searches~\cite{Dolinski:2019nrj,GERDA:2019ivs}.

\begin{figure}[htbp]
    \centering
\includegraphics[height=8cm,width=0.8\linewidth]{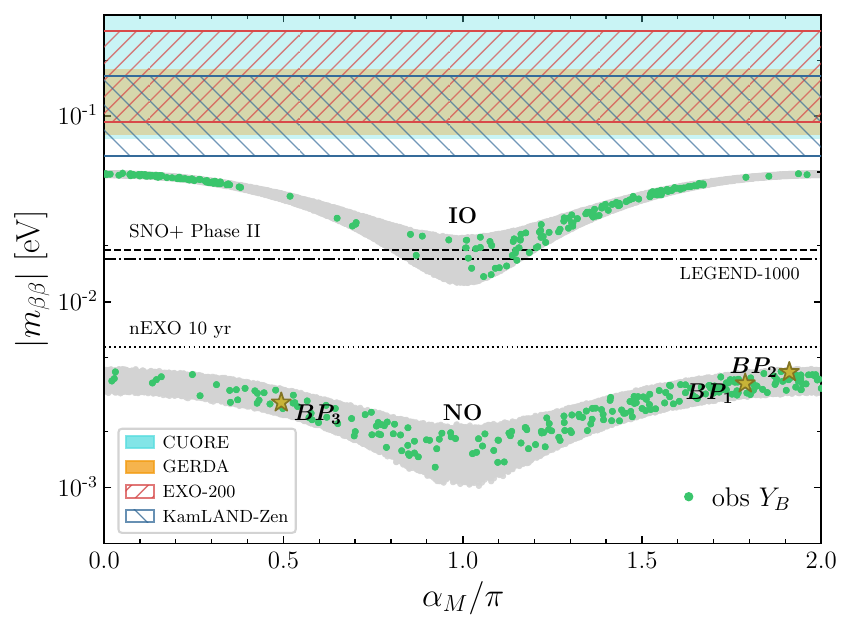}
    \caption{Effective mass parameter characterising the \znbb amplitude versus the relevant Majorana phase, for normal and inverted neutrino spectra. Green points yield viable values of the cosmological baryon asymmetry.}
    \label{fig:n0bb}
\end{figure}

An implication of our minimal linear seesaw scenario is the fact that the lightest neutrino is massless, thanks to the missing partner nature of the seesaw mechanism~\cite{Schechter:1980gr}. For this scenario, Eq.~\eqref{eq:mbb} reduces to the form
\begin{equation}
\label{eq:n0bb_minimal}
\begin{aligned}
    |m_{\beta\beta}|=
    \begin{cases}
        |m_2s_{12}^2c_{13}^2+m_3s_{13}^2e^{i\alpha_M}| & \text{for NO}\\
        c_{13}^2|m_1c_{12}^2+m_2s_{12}^2e^{i\alpha_M}| & \text{for IO},
    \end{cases}
\end{aligned}
\end{equation}
where the Majorana phase $\alpha_M$ is defined as
\begin{equation}
\begin{aligned}
    \alpha_M=\begin{cases}
        -2(\phi_{13}-\phi_{12}) & \text{for NO}\\
        -2\phi_{12} & \text{for IO}.
    \end{cases}
\end{aligned}
\end{equation}
The resulting lower bounds for the \znbb amplitudes of Eq.~\eqref{eq:n0bb_minimal} can be seen in Fig.~\ref{fig:n0bb}. The distribution of successful points illustrates the extent to which the Majorana phase entering neutrinoless double-beta decay correlates with the CP parameter relevant for leptogenesis, see Eq.~\eqref{eq:cp_invariant}. The figure suggests the possibility of probing successful leptogenesis scenarios with \znbb searches at future experimental facilities. 

\begin{figure}[htbp]
    \centering
    \subfigure[]{
        \centering
        \includegraphics[width=0.6\linewidth]{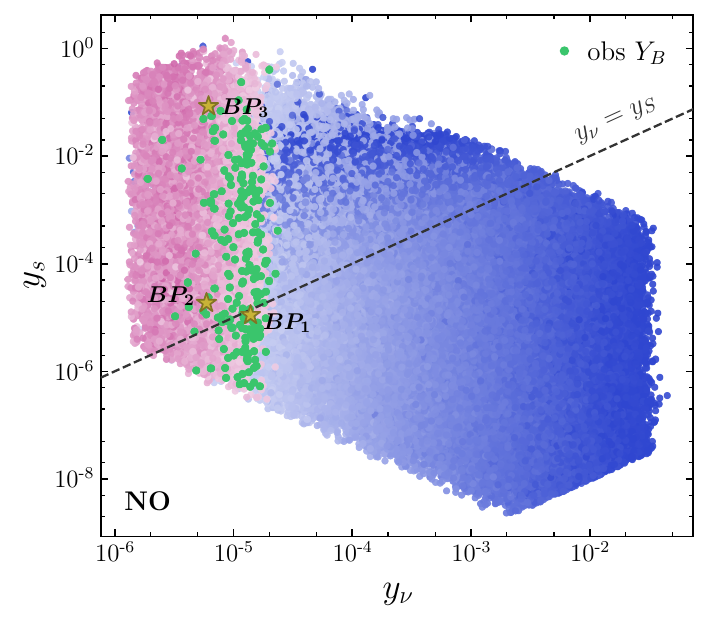}
    }  
    \caption{Cosmological baryon asymmetry versus $y_\nu$ and $y_S$ obtained in the global scan. Colours indicate $Y_{\Delta B}/Y_{\Delta B}^{\rm obs}$, green points lie in the allowed range, and stars mark the benchmarks: BP${}_1$ and BP${}_3$ lie in the Higgs regime while BP${}_2$ is in the leptophilic regime.}
    \label{fig:ynu_ys}
\end{figure}
A remarkable feature of our leptogenesis scenario is that not only do all the relevant new-physics scales lie in the TeV range, but also lepton-number violation is intrinsically associated with a low scale, \(v_\chi\). This allows for sizable values of the Yukawa coupling combinations \(y_\nu\) and \(y_S\), without compromising the smallness of the light-neutrino masses.
The general profile in the $y_{\nu}-y_S$ plane is shown in Fig.~\ref{fig:ynu_ys}.
This figure shows the projection of the parameters obtained in our leptogenesis scan onto the $(y_\nu,y_S)$ plane for a normal-ordered neutrino mass spectrum. 

In the seesaw limit, the mixing between active and heavy neutrinos is controlled predominantly by the matrix $\mathbf{S}$ defined in Eq.~\eqref{eq:S_mixing_matrix}. Consequently, the non-unitarity effects and the heavy-neutrino contributions to charged-lepton flavour violation mediated by the electroweak gauge bosons, such as \(\mu \to e\gamma\), involve combinations such as
$$
(\mathbf{SS}^\dagger)_{e\mu}
\simeq
\frac{v_\Phi^2}{2M_R^2}
(\mathbf{Y}_\nu 
\mathbf{Y}_\nu^\dagger)_{e\mu}.
$$
For fixed $M_R$, these contributions are therefore sensitive to $y_\nu^2$, rather than directly to the product $y_\nu y_S$ which determines neutrino masses \cite{Batra:2023mds,Mandal:2026wbz}.
 The additional charged scalar $H^\pm$ also gives a contribution involving the Yukawa vector $\mathbf{Y}_S$ in the small-$v_\chi$ limit. 

Likewise, the same pattern of Yukawa couplings and scalar masses also governs the production and decays of the heavy-neutrino mediators. In previous works \cite{Batra:2023ssq,Batra:2023mds} the following production channels involving heavy neutrinos and scalars
 \begin{equation}
 e^+e^-\to N_aN_b,\qquad
 e^-\gamma\to N_aH^-,
 \qquad
 e^+e^-\to H^+H^-
 \end{equation}
have been proposed. 
The associated decay chains can lead to multilepton final states, including distinctive signatures with lepton-number violation. 

In short, our scenario can harbour rich, testable phenomenology, with potentially sizable rates for a variety of charged-lepton flavour-violating processes, such as $\mu\to e \gamma$, alongside distinctive signatures at high-energy colliders.
The latter include the remarkable possibility of observing striking signals of lepton-number violation at high-energy colliders~\cite{Batra:2023mds,Batra:2023ssq}, such as the same-sign dilepton plus four-jet signal at the proposed  FCC-${ee}$ collider~\cite{Bharadwaj:2026edq}.

\section{conclusions
}\label{sec:conclusions}

We have shown that the linear seesaw mechanism provides a natural framework for realizing resonant leptogenesis at the TeV scale. Our analysis has focused on the simplest effective scenario, in which neutrino masses are generated through a single pair of heavy neutrino mediators forming a quasi-Dirac state. 
The small lepton-number violation scale suppresses the light neutrino masses and determines the tiny mass splitting of the quasi-Dirac pair. 
Leptogenesis is intrinsically flavoured, since the unflavoured CP-asymmetry cancels in Eq.~\eqref{eq:flavoured-cp-cancellation}, so the final asymmetry arises from differences in washout per flavour.

The resulting lepton asymmetry arises from heavy-neutrino decays (Fig.~\ref{fig:cp_decays}), with the efficiency factors computed in the strong-washout approximation. Our main results are summarized in  Fig.~\ref{fig:K_res}.
As seen, viable values of the observed cosmological baryon asymmetry can be obtained within the resonant leptogenesis regime. Moreover, we found that successful solutions also exist within the regime where heavy neutrinos decay predominantly into the leptophilic scalar rather than the SM Higgs boson.
Viable cosmological asymmetry values are related to low-energy neutrino parameters.
For illustration, the expected values of the neutrinoless double beta decay amplitude and the baryon asymmetry are shown in Fig.~\ref{fig:n0bb}. Furthermore, the pattern of Yukawa parameters displayed in Fig.~\ref{fig:ynu_ys} governs a broad range of complementary phenomenological probes, including potentially striking signatures of charged-lepton flavour violation and lepton-number violation at colliders. A dedicated analysis of the interplay between our leptogenesis framework and these complementary probes lies beyond the scope of the present work and will be pursued elsewhere.

\section*{Acknowledgments}

This work is funded by Spanish grants PID2023-147306NB-I00 and by Severo Ochoa Excellence grant 
CEX2023-001292-S funded by MICIU/AEI. AECH has received funding from Chilean grants ANID-Chile FONDECYT 1261103, 1241855, ANID CCTVal CIA250027 and ANID – Millennium Science Initiative Program code ICN2019$\_$044. SFK acknowledges financial support from the STFC
Consolidated Grant ST/X000583/1. SFK was funded by
a Leverhulme Trust Emeritus Fellowship Grant. J. E. P. was supported by ANID (Chile) under FONDECYT Grant No. 3260491.

\appendix

\section{Decay widths and CP-asymmetry}
\label{app:decay_cp}

Expressing the neutrino Yukawa interactions 
of Eq.~\eqref{eq:yukawa_lagrangian} in terms of physical mass eigenstates defined in Eqs.~\eqref{eq:physical_higgs} and \eqref{eq:physical_leptophilic} and Eq.~\eqref{eq:neutrino_mixing_small}, we obtain 
\begin{equation}
\begin{aligned}
-\mathcal L
={}&
\frac{1}{\sqrt{2}}\sum_{a=4,5}\sum_{i=e,\mu,\tau}
\nu_{Li}^{T}C P_LN_a
\left[
\lambda_{ia}^{(h)}h
+
\lambda_{ia}^{(H)}H
+
\lambda_{ia}^{(A)}A
\right]\\
&+
\sum_{a=4,5}\sum_{i=e,\mu,\tau}
\lambda_{ia}^{(H^\pm)}
\ell_{Li}^{T}C P_LN_a H^+
+
\mathrm{h.c.}
\end{aligned}
\label{eq:lagrangian}
\end{equation}
Here, $i=1,2,3$ denote the lepton flavour and $a=4,5$ denote heavy neutrino. The explicit form of the couplings is 
\begin{equation}
\begin{aligned}
\lambda_{i4}^{(h)}
&=
-\frac{i}{2}
\left(c_\alpha Y_{\nu i}+s_\alpha Y_{Si}\right),
&
\lambda_{i5}^{(h)}
&=
\frac{1}{2}
\left(c_\alpha Y_{\nu i}-s_\alpha Y_{Si}\right),
\\
\lambda_{i4}^{(H)}
&=
-\frac{i}{2}
\left(s_\alpha Y_{\nu i}-c_\alpha Y_{Si}\right),
&
\lambda_{i5}^{(H)}
&=
\frac{1}{2}
\left(s_\alpha Y_{\nu i}+c_\alpha Y_{Si}\right),
\\
\lambda_{i4}^{(A)}
&=
\frac{1}{2}
\left(c_\beta Y_{\nu i}+s_\beta Y_{Si}\right),
&
\lambda_{i5}^{(A)}
&=
\frac{i}{2}
\left(c_\beta Y_{\nu i}-s_\beta Y_{Si}\right),
\\
\lambda_{i4}^{(H^\pm)}
&=
\frac{i}{\sqrt{2}}
\left(c_\beta Y_{\nu i}+s_\beta Y_{Si}\right),
&
\lambda_{i5}^{(H^\pm)}
&=
-\frac{1}{\sqrt{2}}
\left(c_\beta Y_{\nu i}-s_\beta Y_{Si}\right).
\end{aligned}
\label{eq:lambda_couplings}
\end{equation}

The CP-averaged partial decay width into a lepton of flavour $i$ and a scalar channel $r$ is
\begin{equation}
\begin{aligned}
\Gamma_{ai}^{(r)}
&\equiv
\Gamma(N_a\to\ell_i\mathcal H^{(r)})
+
\Gamma(N_a\to\overline{\ell}_i\mathcal H^{(r)*})
=
\frac{M_a}{32\pi}|\lambda_{ia}^{(r)}|^2P_{ar}
\end{aligned}
\label{eq:partial-width-leptogenesis}
\end{equation}
where 
\begin{equation}
P_{ar}
=
\left(1-\frac{m_r^2}{M_a^2}\right)^2
\Theta(M_a-m_r)
\end{equation}
is the kinematical factor of the process. 

We define the leptonic CP-asymmetry of each channel as
\begin{equation}
\begin{aligned}
    \epsilon_{ai}^{(r)}
    &\equiv
    \frac{
    \Gamma(N_a\to\ell_i\mathcal H^{(r)})
    -
    \Gamma(N_a\to\overline{\ell}_i\mathcal H^{(r)*})
    }{\Gamma_a}.
\end{aligned}
\end{equation}
In the resonant leptogenesis regime, this contribution is given by \cite{Fu:2021fyk} 
\begin{equation}
\label{eq:resonant_cp_asymmetry}
\epsilon_{ai}^{(r)}
=-
\frac{
P_{ar}\operatorname{Im}\!\left[
\lambda_{i4}^{(r)}\lambda_{i5}^{(r)*}
\sum_{s,j}P_{as}\lambda_{j4}^{(s)}\lambda_{j5}^{(s)*}
\right]
}{
\left[\sum_{s,j}P_{4s}|\lambda_{j4}^{(s)}|^2\right]
\left[\sum_{s,j}P_{5s}|\lambda_{j5}^{(s)}|^2\right]
}
\frac{2\Delta M_N\Gamma_b}{4(\Delta M_N)^2+\Gamma_b^2},
\end{equation}
with $(a,b)=(4,5),(5,4)$ and $r,s=h,H,A,H^\pm$. 

In this model the absorptive sum is purely imaginary. As a consequence, the numerator the equation above can be factorized as 
\begin{equation}
\begin{aligned}
\operatorname{Im}\!\left[
\lambda_{i4}^{(r)}\lambda_{i5}^{(r)*}
\sum_{s,j}P_{as}\lambda_{j4}^{(s)}\lambda_{j5}^{(s)*}
\right]
={}&
\operatorname{Re}\!\left[
\lambda_{i4}^{(r)}\lambda_{i5}^{(r)*}
\right]
\operatorname{Im}\!\left[
\sum_{s,j}P_{as}\lambda_{j4}^{(s)}\lambda_{j5}^{(s)*}
\right]=\eta_i^{\mathrm{CP}}\kappa_a^{(r)}.
\end{aligned}
\label{eq:factorization}
\end{equation}
Using Eq.~\eqref{eq:lambda_couplings} 
\begin{equation}
\begin{aligned}
&\kappa_a^{(r)}
=
\frac{
\operatorname{Im}\!\left[
\displaystyle\sum_{s,j}P_{as}
\lambda_{j4}^{(s)}\lambda_{j5}^{(s)*}
\right]
}{
2\left[\displaystyle\sum_{s,j}P_{4s}|\lambda_{j4}^{(s)}|^2\right]
\left[\displaystyle\sum_{s,j}P_{5s}|\lambda_{j5}^{(s)}|^2\right]
}
\begin{cases}
P_{ah}\sin2\alpha, & r=h,\\
-P_{aH}\sin2\alpha, & r=H,\\
P_{aA}\sin2\beta, & r=A,\\
2P_{aH^\pm}\sin2\beta, & r=H^\pm
\end{cases}\\
&\eta_i^{\mathrm{CP}}
=\operatorname{Im}(Y_{Si}^{*}Y_{\nu i}).
\end{aligned}
\label{eq:kappa_and_eta}
\end{equation}
In the approximation $M_4\simeq M_5\simeq M_R$, $P_{4r}\simeq P_{5r}\simeq P_r$, the factor $\kappa^{(r)}$ results in
\begin{equation}
\begin{aligned}
\kappa^{(r)}
\simeq{}&
-\frac{1}{y_\nu y_S}
\Bigg[
\frac{
P_hc_\alpha^2+P_Hs_\alpha^2
+(P_A+2P_{H^\pm})c_\beta^2
}{r_y}
-r_y
\left(
P_hs_\alpha^2+P_Hc_\alpha^2
+(P_A+2P_{H^\pm})s_\beta^2
\right)
\Bigg]
\\[1mm]
&\times
\Bigg\{
\Bigg[
\frac{
P_hc_\alpha^2+P_Hs_\alpha^2
+(P_A+2P_{H^\pm})c_\beta^2
}{r_y}
+r_y
\left(
P_hs_\alpha^2+P_Hc_\alpha^2
+(P_A+2P_{H^\pm})s_\beta^2
\right)
\Bigg]^2
\\
&\hspace{6mm}
-\rho^2
\left[
(P_h-P_H)\sin2\alpha
+(P_A+2P_{H^\pm})\sin2\beta
\right]^2
\Bigg\}^{-1}
\begin{cases}
P_h\sin2\alpha, & r=h,\\
-P_H\sin2\alpha, & r=H,\\
P_A\sin2\beta, & r=A,\\
2P_{H^\pm}\sin2\beta, & r=H^\pm.
\end{cases}
\end{aligned}
\label{eq:kappa_full}
\end{equation}

\section{Details on quasi-Dirac heavy pair}

The most general neutral fermion mass matrix is Majorana, hence complex and symmetric, so that the physical mass basis is defined by a complex diagonalization~\cite{Schechter:1980gr}. In our minimal  linear seesaw mechanism this leads to 
\begin{equation}
\mathcal{U}_0^T\mathcal{M}_\nu\mathcal{U}_0
=
D,
\qquad
D
=
\operatorname{diag}(m_1,m_2,m_3,M_4,M_5).
\end{equation}
In the seesaw regime the diagonalization can be organized in two steps~\cite{Schechter:1981cv},
\begin{equation}
\mathcal{U}_0=\mathcal{W}\mathcal{V},
\end{equation}
where $\mathcal{W}$ block-diagonalizes the full matrix, while $\mathcal{V}$ diagonalizes the light and heavy blocks separately: 
\begin{equation}
\mathcal{W}^T\mathcal{M}_\nu\mathcal{W}
=
\begin{pmatrix}
m_\nu & 0\\
0 & M_H
\end{pmatrix},
\qquad
\mathcal{V}
=
\begin{pmatrix}
K & 0\\
0 & U_H
\end{pmatrix}.
\end{equation}
Thus we set, for the full matrix,  
\begin{equation}
K^T m_\nu K=D_\nu,
\qquad
U_H^TM_H U_H=D_H,
\end{equation}
with
\begin{equation}
D_\nu=\operatorname{diag}(m_1,m_2,m_3),
\qquad
D_H=\operatorname{diag}(M_4,M_5)
\end{equation}
so that all mass eigenvalues will be positive.  
In the lepton number conserving limit, the heavy block is simply the Dirac mass term for the mediator, namely, 
\begin{equation}
M_H^{(0)}
=
\begin{pmatrix}
0 & M_R\\
M_R & 0
\end{pmatrix}.
\end{equation}
A real rotation gives 
\begin{equation}
O_H^TM_H^{(0)}O_H
=
\operatorname{diag}(-M_R,M_R),
\qquad
O_H=
\frac{1}{\sqrt2}
\begin{pmatrix}
1 & 1\\
-1 & 1
\end{pmatrix}.
\end{equation}
The block is diagonal, but one mass eigenvalue is negative. We can define 
\begin{equation}
P_\pm=\operatorname{diag}(\pm i,1) 
\end{equation}
so that 
\begin{equation}
P_\pm^T\operatorname{diag}(-M_R,M_R)P_\pm
=
\operatorname{diag}(M_R,M_R),
\end{equation}
and $U_H=O_H P_\pm$. The factor $\pm i$ converts the Majorana eigenvalue into a positive mass~\cite{Schechter:1980gr}. Both signs differ only by a redefinition of the Majorana field. 

\black
\bibliographystyle{utphys}
\bibliography{bibliography} 

@article{Buchmuller:2005eh,
    author = "Buchmuller, W. and Peccei, R. D. and Yanagida, T.",
    title = "{Leptogenesis as the origin of matter}",
    eprint = "hep-ph/0502169",
    archivePrefix = "arXiv",
    reportNumber = "DESY-05-031",
    doi = "10.1146/annurev.nucl.55.090704.151558",
    journal = "Ann. Rev. Nucl. Part. Sci.",
    volume = "55",
    pages = "311--355",
    year = "2005"
}

@article{Chikashige:1980ui,
    author = "Chikashige, Y. and Mohapatra, Rabindra N. and Peccei, R. D.",
    title = "{Are There Real Goldstone Bosons Associated with Broken Lepton Number?}",
    reportNumber = "MPI-PAE-PTH-36-80",
    doi = "10.1016/0370-2693(81)90011-3",
    journal = "Phys. Lett. B",
    volume = "98",
    pages = "265--268",
    year = "1981"
}

@article{Chakraborty:2021azg,
    author = "Chakraborty, Indrani and Roy, Himadri and Srivastava, Tripurari",
    title = "{Resonant leptogenesis in (2,2) inverse see-saw realisation}",
    eprint = "2106.08232",
    archivePrefix = "arXiv",
    primaryClass = "hep-ph",
    doi = "10.1016/j.nuclphysb.2022.115780",
    journal = "Nucl. Phys. B",
    volume = "979",
    pages = "115780",
    year = "2022"
}

@article{Chianese:2021toe,
    author = "Chianese, Marco and Fu, Bowen and King, Stephen F.",
    title = "{Dark Matter in the Type Ib Seesaw Model}",
    eprint = "2102.07780",
    archivePrefix = "arXiv",
    primaryClass = "hep-ph",
    doi = "10.1007/JHEP05(2021)129",
    journal = "JHEP",
    volume = "05",
    pages = "129",
    year = "2021"
}

@article{Bharadwaj:2026edq,
    author = "Bharadwaj, Praveen and Mandal, Sanjoy and Padhan, Rojalin and Valle, Jos{\'e} W. F.",
    title = "{Probing lepton number violation at FCC-ee}",
    eprint = "2606.09993",
    archivePrefix = "arXiv",
    primaryClass = "hep-ph",
    month = "6",
    year = "2026"
}

@article{Hernandez-Garcia:2019uof,
    author = "Hernandez-Garcia, Josu and King, Stephen F.",
    title = "{New Weinberg operator for neutrino mass and its seesaw origin}",
    eprint = "1903.01474",
    archivePrefix = "arXiv",
    primaryClass = "hep-ph",
    doi = "10.1007/JHEP05(2019)169",
    journal = "JHEP",
    volume = "05",
    pages = "169",
    year = "2019"
}

@article{Fu:2021fyk,
    author = "Fu, Bowen and King, Stephen F.",
    title = "{Leptogenesis in type Ib seesaw models}",
    eprint = "2107.01486",
    archivePrefix = "arXiv",
    primaryClass = "hep-ph",
    doi = "10.1103/PhysRevD.105.095001",
    journal = "Phys. Rev. D",
    volume = "105",
    number = "9",
    pages = "095001",
    year = "2022"
}

@article{Fukugita:1986hr,
    author = "Fukugita, M. and Yanagida, T.",
    title = "{Baryogenesis Without Grand Unification}",
    reportNumber = "RIFP-641",
    doi = "10.1016/0370-2693(86)91126-3",
    journal = "Phys. Lett. B",
    volume = "174",
    pages = "45--47",
    year = "1986"
}

@article{Pilaftsis:2003gt,
    author = "Pilaftsis, Apostolos and Underwood, Thomas E. J.",
    title = "{Resonant leptogenesis}",
    eprint = "hep-ph/0309342",
    archivePrefix = "arXiv",
    reportNumber = "MC-TH-2003-09",
    doi = "10.1016/j.nuclphysb.2004.05.029",
    journal = "Nucl. Phys. B",
    volume = "692",
    pages = "303--345",
    year = "2004"
}

@article{Dolinski:2019nrj,
    author = "Dolinski, Michelle J. and Poon, Alan W. P. and Rodejohann, Werner",
    title = "{Neutrinoless Double-Beta Decay: Status and Prospects}",
    eprint = "1902.04097",
    archivePrefix = "arXiv",
    primaryClass = "nucl-ex",
    doi = "10.1146/annurev-nucl-101918-023407",
    journal = "Ann. Rev. Nucl. Part. Sci.",
    volume = "69",
    pages = "219--251",
    year = "2019"
}

@article{Planck:2018vyg,
    author = "Aghanim, N. and others",
    collaboration = "Planck",
    title = "{Planck 2018 results. VI. Cosmological parameters}",
    eprint = "1807.06209",
    archivePrefix = "arXiv",
    primaryClass = "astro-ph.CO",
    doi = "10.1051/0004-6361/201833910",
    journal = "Astron. Astrophys.",
    volume = "641",
    pages = "A6",
    year = "2020",
    note = "[Erratum: Astron.Astrophys. 652, C4 (2021)]"
}

@article{Rodejohann:2011vc,
    author = "Rodejohann, W. and Valle, J. W. F.",
    title = "{Symmetrical Parametrizations of the Lepton Mixing Matrix}",
    eprint = "1108.3484",
    archivePrefix = "arXiv",
    primaryClass = "hep-ph",
    reportNumber = "IFIC-11-39",
    doi = "10.1103/PhysRevD.84.073011",
    journal = "Phys. Rev. D",
    volume = "84",
    pages = "073011",
    year = "2011"
}

@article{Schechter:1980gr,
    author = "Schechter, J. and Valle, J. W. F.",
    title = "{Neutrino Masses in SU(2) x U(1) Theories}",
    reportNumber = "SU-4217-167, COO-3533-167",
    doi = "10.1103/PhysRevD.22.2227",
    journal = "Phys. Rev. D",
    volume = "22",
    pages = "2227",
    year = "1980"
}

@article{Fontes:2019uld,
    author = "Fontes, Duarte and Romao, Jorge C. and Valle, Jose W. F.",
    title = "{Electroweak Breaking and Higgs Boson Profile in the Simplest Linear Seesaw Model}",
    eprint = "1908.09587",
    archivePrefix = "arXiv",
    primaryClass = "hep-ph",
    reportNumber = "IFIC/19-XX, CFTP/19-024",
    doi = "10.1007/JHEP10(2019)245",
    journal = "JHEP",
    volume = "10",
    pages = "245",
    year = "2019"
}

@article{Akhmedov:1995ip,
    author = "Akhmedov, Evgeny K. and others",
    title = "{Left-right symmetry breaking in NJL approach}",
    eprint = "hep-ph/9507275",
    archivePrefix = "arXiv",
    reportNumber = "IC-95-125, TUM-HEP-221-95, MPI-PHT-95-35, FTUV-95-34, IFIC-95-36",
    doi = "10.1016/0370-2693(95)01504-3",
    journal = "Phys. Lett. B",
    volume = "368",
    pages = "270--280",
    year = "1996"
}

@article{Akhmedov:1995vm,
    author = "Akhmedov, Evgeny K. and others",
    title = "{Dynamical left-right symmetry breaking}",
    eprint = "hep-ph/9509255",
    archivePrefix = "arXiv",
    reportNumber = "IC-95-126, TUM-HEP-222-95, MPI-PHT-95-70, FTUV-95-36, IFIC-95-38",
    doi = "10.1103/PhysRevD.53.2752",
    journal = "Phys. Rev. D",
    volume = "53",
    pages = "2752--2780",
    year = "1996"
}

@article{Malinsky:2005bi,
    author = "Malinsky, Michal and Romao, J. C. and Valle, J. W. F.",
    title = "{Novel supersymmetric SO(10) seesaw mechanism}",
    eprint = "hep-ph/0506296",
    archivePrefix = "arXiv",
    reportNumber = "IFIC-05-28",
    doi = "10.1103/PhysRevLett.95.161801",
    journal = "Phys. Rev. Lett.",
    volume = "95",
    pages = "161801",
    year = "2005"
}

@article{deSalas:2020pgw,
    author = "de Salas, P. F. and others",
    title = "{2020 global reassessment of the neutrino oscillation picture}",
    eprint = "2006.11237",
    archivePrefix = "arXiv",
    primaryClass = "hep-ph",
    doi = "10.1007/JHEP02(2021)071",
    journal = "JHEP",
    volume = "02",
    pages = "071",
    year = "2021"
}

@article{Schechter:1981cv,
    author = "Schechter, J. and Valle, J. W. F.",
    title = "{Neutrino Decay and Spontaneous Violation of Lepton Number}",
    reportNumber = "SU-4217-203, COO-3533-203",
    doi = "10.1103/PhysRevD.25.774",
    journal = "Phys. Rev. D",
    volume = "25",
    pages = "774",
    year = "1982"
}

@article{Batra:2022arl,
    author = "Batra, Aditya and others",
    title = "{W-mass anomaly in the simplest linear seesaw mechanism}",
    eprint = "2208.04983",
    archivePrefix = "arXiv",
    primaryClass = "hep-ph",
    reportNumber = "IFIC/22-XXX",
    doi = "10.1016/j.physletb.2022.137408",
    journal = "Phys. Lett. B",
    volume = "834",
    pages = "137408",
    year = "2022"
}

@Article{GERDA:2019ivs,
        author = "Agostini, M. and others",
 collaboration = "GERDA",
         title = "{Probing Majorana neutrinos with double-$\beta$ decay}",
       journal = "Science",
        volume = "365",
          year = "2019",
         pages = "1445",
 archiveprefix = "arXiv",
  primaryclass = "hep-ex",
        eprint = "1909.02726",
           doi = "10.1126/science.aav8613",
}

@Article{Schechter:1981bd,
        author = "Schechter, J. and Valle, J.~W.~F.",
         title = "{Neutrinoless Double beta Decay in SU(2) x U(1) Theories}",
       journal = "Phys.Rev.D",
        volume = "25",
          year = "1982",
         pages = "2951",
           doi = "10.1103/PhysRevD.25.2951",
  reportnumber = "SU-4217-213, COO-3533-213",
}

@article{Batra:2023mds,
    author = "Batra, Aditya and others",
    title = "{Phenomenology of the simplest linear seesaw mechanism}",
    eprint = "2305.00994",
    archivePrefix = "arXiv",
    primaryClass = "hep-ph",
    doi = "10.1007/JHEP07(2023)221",
    journal = "JHEP",
    volume = "07",
    pages = "221",
    year = "2023"
}

@article{Batra:2023ssq,
    author = "Batra, Aditya and others",
    title = "{Large lepton number violation at colliders: Predictions from the minimal linear seesaw mechanism}",
    eprint = "2304.06080",
    archivePrefix = "arXiv",
    primaryClass = "hep-ph",
    doi = "10.1016/j.physletb.2024.139204",
    journal = "Phys. Lett. B",
    volume = "860",
    pages = "139204",
    year = "2025"
}

@article{Davidson:2008bu,
    author = "Davidson, Sacha and Nardi, Enrico and Nir, Yosef",
    title = "{Leptogenesis}",
    eprint = "0802.2962",
    archivePrefix = "arXiv",
    primaryClass = "hep-ph",
    doi = "10.1016/j.physrep.2008.06.002",
    journal = "Phys. Rept.",
    volume = "466",
    pages = "105--177",
    year = "2008"
}

@article{Eijima:2017anv,
    author = "Eijima, Shintaro and Shaposhnikov, Mikhail",
    title = "{Fermion number violating effects in low scale leptogenesis}",
    eprint = "1703.06085",
    archivePrefix = "arXiv",
    primaryClass = "hep-ph",
    doi = "10.1016/j.physletb.2017.05.068",
    journal = "Phys. Lett. B",
    volume = "771",
    pages = "288--296",
    year = "2017"
}

@article{Mandal:2026wbz,
    author = "Mandal, Sanjoy and Roshan, Rishav and Valle, Jose W. F.",
    title = "{Particle and gravitational wave probes of minimal seesaw neutrinos}",
    eprint = "2606.15316",
    archivePrefix = "arXiv",
    primaryClass = "hep-ph",
    doi = "10.1007/JHEP09(2026)173",
    journal = "JHEP",
    volume = "09",
    pages = "173",
    year = "2026"
}

@article{Buchmuller:2004nz,
    author = "Buchmuller, W. and Di Bari, P. and Plumacher, M.",
    title = "{Leptogenesis for pedestrians}",
    eprint = "hep-ph/0401240",
    archivePrefix = "arXiv",
    reportNumber = "DESY-03-100, UAB-FT-551, CERN-TH-2003-199",
    doi = "10.1016/j.aop.2004.02.003",
    journal = "Annals Phys.",
    volume = "315",
    pages = "305--351",
    year = "2005"
}

@article{CMS:2026zxu,
    author = "Hayrapetyan, Aram and others",
    collaboration = "CMS",
    title = "{Combined measurements and interpretations of Higgs boson production and decay in proton{\textendash}proton collisions at $\sqrt{s} = 13\,\textrm{TeV}$}",
    eprint = "2602.18611",
    archivePrefix = "arXiv",
    primaryClass = "hep-ex",
    reportNumber = "CMS-HIG-21-018, CERN-EP-2026-009",
    doi = "10.1088/1361-6633/ae72c4",
    journal = "Rept. Prog. Phys.",
    volume = "89",
    number = "8",
    pages = "087801",
    year = "2026"
}

@article{ATLAS:2026pdi,
    author = "Aad, Georges and others",
    collaboration = "ATLAS",
    title = "{Combination of Higgs boson measurements at $\sqrt{s} =$ 13 TeV and their interpretations by the ATLAS experiment}",
    eprint = "2608.07332",
    archivePrefix = "arXiv",
    primaryClass = "hep-ex",
    reportNumber = "CERN-EP-2026-228",
    month = "8",
    year = "2026"
}

@article{Dolan:2018yqy,
    author = "Dolan, Matthew J. and Dutka, Tomasz P. and Volkas, Raymond R.",
    title = "{Low-scale leptogenesis with minimal lepton flavor violation}",
    eprint = "1812.11964",
    archivePrefix = "arXiv",
    primaryClass = "hep-ph",
    doi = "10.1103/PhysRevD.99.123508",
    journal = "Phys. Rev. D",
    volume = "99",
    number = "12",
    pages = "123508",
    year = "2019"
}

@article{Abada:2023zbb,
    author = "Abada, Asmaa and Bernal, Nicol{\'a}s and C{\'a}rcamo Hern{\'a}ndez, Antonio E. and Kovalenko, Sergey and de Melo, T{\'e}ssio B.",
    title = "{Three-loop inverse scotogenic seesaw models}",
    eprint = "2312.14105",
    archivePrefix = "arXiv",
    primaryClass = "hep-ph",
    doi = "10.1007/JHEP05(2024)035",
    journal = "JHEP",
    volume = "05",
    pages = "035",
    year = "2024"
}

@article{Abada:2026ons,
    author = "Abada, Asmaa and Bernal, Nicol{\'a}s and C{\'a}rcamo Hern{\'a}ndez, A. E. and Vishnudath, K. N. and Kovalenko, Sergey and de Melo, T{\'e}ssio B. and Urrea, Salvador",
    title = "{Multi-component Dark Matter in a Novel Three-Loop Inverse Scotogenic Seesaw Model}",
    eprint = "2607.24932",
    archivePrefix = "arXiv",
    primaryClass = "hep-ph",
    month = "7",
    year = "2026"
}

@article{Deppisch:2010fr,
    author = "Deppisch, Frank F. and Pilaftsis, Apostolos",
    title = "{Lepton Flavour Violation and theta(13) in Minimal Resonant Leptogenesis}",
    eprint = "1012.1834",
    archivePrefix = "arXiv",
    primaryClass = "hep-ph",
    reportNumber = "MAN-HEP-2010-22",
    doi = "10.1103/PhysRevD.83.076007",
    journal = "Phys. Rev. D",
    volume = "83",
    pages = "076007",
    year = "2011"
}

@article{BhupalDev:2014pfm,
    author = "Bhupal Dev, P. S. and Millington, Peter and Pilaftsis, Apostolos and Teresi, Daniele",
    title = "{Flavour Covariant Transport Equations: an Application to Resonant Leptogenesis}",
    eprint = "1404.1003",
    archivePrefix = "arXiv",
    primaryClass = "hep-ph",
    reportNumber = "MAN-HEP-2014-01, IPPP-14-20, DCPT-14-40",
    doi = "10.1016/j.nuclphysb.2014.06.020",
    journal = "Nucl. Phys. B",
    volume = "886",
    pages = "569--664",
    year = "2014"
}

@article{ParticleDataGroup:2026mpi,
    author = "Takahashi, F. and others",
    collaboration = "Particle Data Group",
    title = "{Review of Particle Physics}",
    doi = "10.1142/s0217751x26300115",
    journal = "Int. J. Mod. Phys. A",
    volume = "41",
    number = "22",
    pages = "2630011",
    year = "2026"
}

@article{Naredo-Tuero:2024sgf,
    author = "Naredo-Tuero, Daniel and Escudero, Miguel and Fern{\'a}ndez-Mart{\'\i}nez, Enrique and Marcano, Xabier and Poulin, Vivian",
    title = "{Critical look at the cosmological neutrino mass bound}",
    eprint = "2407.13831",
    archivePrefix = "arXiv",
    primaryClass = "astro-ph.CO",
    reportNumber = "CERN-TH-2024-115, IFT-UAM/CSIC-24-106",
    doi = "10.1103/PhysRevD.110.123537",
    journal = "Phys. Rev. D",
    volume = "110",
    number = "12",
    pages = "123537",
    year = "2024"
}

@article{DESI:2025zgx,
    author = "Abdul Karim, M. and others",
    collaboration = "DESI",
    title = "{DESI DR2 results. II. Measurements of baryon acoustic oscillations and cosmological constraints}",
    eprint = "2503.14738",
    archivePrefix = "arXiv",
    primaryClass = "astro-ph.CO",
    reportNumber = "FERMILAB-PUB-25-0169-PPD",
    doi = "10.1103/tr6y-kpc6",
    journal = "Phys. Rev. D",
    volume = "112",
    number = "8",
    pages = "083515",
    year = "2025"
}

@article{Jiang:2024viw,
    author = "Jiang, Jun-Qian and Giar{\`e}, William and Gariazzo, Stefano and Dainotti, Maria Giovanna and Di Valentino, Eleonora and Mena, Olga and Pedrotti, Davide and da Costa, Simony Santos and Vagnozzi, Sunny",
    title = "{Neutrino cosmology after DESI: tightest mass upper limits, preference for the normal ordering, and tension with terrestrial observations}",
    eprint = "2407.18047",
    archivePrefix = "arXiv",
    primaryClass = "astro-ph.CO",
    doi = "10.1088/1475-7516/2025/01/153",
    journal = "JCAP",
    volume = "01",
    pages = "153",
    year = "2025"
}

@article{Blanchet:2009kk,
    author = "Blanchet, Steve and Hambye, Thomas and Josse-Michaux, Francois-Xavier",
    title = "{Reconciling leptogenesis with observable mu ---{\ensuremath{>}} e gamma rates}",
    eprint = "0912.3153",
    archivePrefix = "arXiv",
    primaryClass = "hep-ph",
    reportNumber = "UMD-PP-09-060, ULB-TH-09-38",
    doi = "10.1007/JHEP04(2010)023",
    journal = "JHEP",
    volume = "04",
    pages = "023",
    year = "2010"
}

@article{Binh:2024lez,
    author = "Binh, V. H. and Bonilla, Cesar and C{\'a}rcamo Hern{\'a}ndez, A. E. and Huong, D. T. and K. N., Vishnudath and Long, H. N. and Thu, P. N. and Schmidt, Iv{\'a}n",
    title = "{Phenomenology of 3-3-1 models with a radiative inverse seesaw mechanism}",
    eprint = "2404.13373",
    archivePrefix = "arXiv",
    primaryClass = "hep-ph",
    doi = "10.1103/PhysRevD.110.075022",
    journal = "Phys. Rev. D",
    volume = "110",
    number = "7",
    pages = "075022",
    year = "2024"
}

@article{Hernandez:2021uxx,
    author = "Hern{\'a}ndez, A. E. C{\'a}rcamo and Schmidt, Ivan",
    title = "{A renormalizable left-right symmetric model with low scale seesaw mechanisms}",
    eprint = "2101.02718",
    archivePrefix = "arXiv",
    primaryClass = "hep-ph",
    doi = "10.1016/j.nuclphysb.2022.115696",
    journal = "Nucl. Phys. B",
    volume = "976",
    pages = "115696",
    year = "2022"
}

@article{Blanchet:2010kw,
    author = "Blanchet, Steve and Dev, P. S. Bhupal and Mohapatra, R. N.",
    title = "{Leptogenesis with TeV Scale Inverse Seesaw in SO(10)}",
    eprint = "1010.1471",
    archivePrefix = "arXiv",
    primaryClass = "hep-ph",
    reportNumber = "UMD-PP-10-016, IFT-UAM-CSIC-10-53, FTUAM-10-14",
    doi = "10.1103/PhysRevD.82.115025",
    journal = "Phys. Rev. D",
    volume = "82",
    pages = "115025",
    year = "2010"
}

@article{Mukherjee:2023nyi,
    author = "Mukherjee, Ananya and Saha, Abhijit Kumar",
    title = "{Rescuing leptogenesis parameter space of inverse seesaw}",
    eprint = "2307.14405",
    archivePrefix = "arXiv",
    primaryClass = "hep-ph",
    doi = "10.1016/j.physletb.2024.138474",
    journal = "Phys. Lett. B",
    volume = "849",
    pages = "138474",
    year = "2024"
}

@article{Hernandez:2021xet,
    author = "Hern{\'a}ndez, A. E. C{\'a}rcamo and Huong, D. T. and Schmidt, Ivan",
    title = "{Universal inverse seesaw mechanism as a source of the SM fermion mass hierarchy}",
    eprint = "2109.12118",
    archivePrefix = "arXiv",
    primaryClass = "hep-ph",
    doi = "10.1140/epjc/s10052-022-10011-x",
    journal = "Eur. Phys. J. C",
    volume = "82",
    number = "1",
    pages = "63",
    year = "2022"
}

@article{Abada:2017ieq,
    author = "Abada, Asmaa and Arcadi, Giorgio and Domcke, Valerie and Lucente, Michele",
    title = "{Neutrino masses, leptogenesis and dark matter from small lepton number violation?}",
    eprint = "1709.00415",
    archivePrefix = "arXiv",
    primaryClass = "hep-ph",
    reportNumber = "DESY-17-124, CP3-17-23, LPT-Orsay-17-35, LPT-ORSAY-17-35",
    doi = "10.1088/1475-7516/2017/12/024",
    journal = "JCAP",
    volume = "12",
    pages = "024",
    year = "2017"
}

@article{Hernandez:2021kju,
    author = "Hern{\'a}ndez, A. E. C{\'a}rcamo and Espinoza, Catalina and G{\'o}mez-Izquierdo, Juan Carlos and Mondrag{\'o}n, Myriam",
    title = "{Fermion masses and mixings, dark matter, leptogenesis and $g-2$ muon anomaly in an extended 2HDM with inverse seesaw}",
    eprint = "2104.02730",
    archivePrefix = "arXiv",
    primaryClass = "hep-ph",
    doi = "10.1140/epjp/s13360-022-03432-w",
    journal = "Eur. Phys. J. Plus",
    volume = "137",
    number = "11",
    pages = "1224",
    year = "2022"
}

@article{Aoki:2015owa,
    author = "Aoki, Mayumi and Haba, Naoyuki and Takahashi, Ryo",
    title = "{A model realizing inverse seesaw and resonant leptogenesis}",
    eprint = "1506.06946",
    archivePrefix = "arXiv",
    primaryClass = "hep-ph",
    reportNumber = "KANAZAWA-15-08, SU-HET-06-2015",
    doi = "10.1093/ptep/ptv149",
    journal = "PTEP",
    volume = "2015",
    number = "11",
    pages = "113B03",
    year = "2015"
}

@article{Agashe:2018cuf,
    author = "Agashe, Kaustubh and Du, Peizhi and Ekhterachian, Majid and Fong, Chee Sheng and Hong, Sungwoo and Vecchi, Luca",
    title = "{Natural Seesaw and Leptogenesis from Hybrid of High-Scale Type I and TeV-Scale Inverse}",
    eprint = "1812.08204",
    archivePrefix = "arXiv",
    primaryClass = "hep-ph",
    doi = "10.1007/JHEP04(2019)029",
    journal = "JHEP",
    volume = "04",
    pages = "029",
    year = "2019"
}

@article{Pilaftsis:2005rv,
    author = "Pilaftsis, Apostolos and Underwood, Thomas E. J.",
    title = "{Electroweak-scale resonant leptogenesis}",
    eprint = "hep-ph/0506107",
    archivePrefix = "arXiv",
    doi = "10.1103/PhysRevD.72.113001",
    journal = "Phys. Rev. D",
    volume = "72",
    pages = "113001",
    year = "2005"
}

@article{Gu:2010xc,
    author = "Gu, Pei-Hong and Sarkar, Utpal",
    title = "{Leptogenesis with Linear, Inverse or Double Seesaw}",
    eprint = "1007.2323",
    archivePrefix = "arXiv",
    primaryClass = "hep-ph",
    doi = "10.1016/j.physletb.2010.09.062",
    journal = "Phys. Lett. B",
    volume = "694",
    pages = "226--232",
    year = "2011"
}

@article{Abada:2015rta,
    author = "Abada, Asmaa and Arcadi, Giorgio and Domcke, Valerie and Lucente, Michele",
    title = "{Lepton number violation as a key to low-scale leptogenesis}",
    eprint = "1507.06215",
    archivePrefix = "arXiv",
    primaryClass = "hep-ph",
    reportNumber = "SISSA-30-2015-FISI, LPT-ORSAY-15-50",
    doi = "10.1088/1475-7516/2015/11/041",
    journal = "JCAP",
    volume = "11",
    pages = "041",
    year = "2015"
}

@article{CarcamoHernandez:2024edi,
    author = "C{\'a}rcamo Hern{\'a}ndez, A. E. and Vel{\'a}squez, Yocelyne Hidalgo and Kovalenko, Sergey and P{\'e}rez-Julve, Nicol{\'a}s A. and Schmidt, Ivan",
    title = "{Models of Radiative Linear Seesaw with Electrically Charged Mediators}",
    eprint = "2403.05637",
    archivePrefix = "arXiv",
    primaryClass = "hep-ph",
    doi = "10.1093/ptep/ptae142",
    journal = "PTEP",
    volume = "2024",
    number = "10",
    pages = "103B02",
    year = "2024"
}

@article{King:2026hnn,
    author = "King, Stephen F. and Manna, Soumen Kumar and Roshan, Rishav and Sil, Arunansu",
    title = "{Leptogenesis with a GeV-Scale Quasi-Dirac Neutrino}",
    eprint = "2610.20729",
    archivePrefix = "arXiv",
    primaryClass = "hep-ph",
    month = "10",
    year = "2026"
}
\end{document}